\documentclass[11pt, a4paper]{article}
\pdfoutput=1
\usepackage{jcappub}

\usepackage{amssymb,amsmath,latexsym,mathrsfs}
\usepackage{graphicx} 
\usepackage{bm}
\usepackage{url}
\usepackage{epsfig,graphicx,verbatim,xspace,multirow,mathtools}
\usepackage{array}
\usepackage{enumitem}

\begin{document}

\hspace*{124mm}{\tt KCL-2018-76}

\title{Neutrino decoupling beyond the Standard Model: \\ CMB constraints on the Dark Matter mass with a fast and precise $N_{\rm eff}$ evaluation
}

\author{Miguel Escudero}
\emailAdd{miguel.escudero@kcl.ac.uk}

\affiliation{Theoretical Particle Physics and Cosmology Group \\
Department of Physics, King's College London, Strand, London WC2R 2LS, UK}

\abstract{The number of effective relativistic neutrino species represents a fundamental probe of the thermal history of the early Universe, and as such of the Standard Model of Particle Physics. Traditional approaches to the process of neutrino decoupling are either very technical and computationally expensive, or assume that neutrinos decouple instantaneously. In this work, we aim to fill the gap between these two approaches by modeling neutrino decoupling in terms of two simple coupled differential equations for the electromagnetic and neutrino sector temperatures, in which all the relevant interactions are taken into account and which allows for a straightforward implementation of BSM species. Upon including finite temperature QED corrections we reach an accuracy on $N_{\rm eff}$ in the SM of $0.01$. We illustrate the usefulness of this approach to neutrino decoupling by considering, in a model independent manner, the impact of MeV thermal dark matter on $N_{\rm eff}$. We show that Planck rules out electrophilic and neutrinophilic thermal dark matter particles of $m< 3.0\,\text{MeV}$ at 95\% CL regardless of their spin, and of their annihilation being $s$-wave or $p$-wave. We point out that thermal dark matter particles with non-negligible interactions with both electrons and neutrinos are more elusive to CMB observations than purely electrophilic or neutrinophilic ones. In addition, assisted by the accuracy of our approach, we show that CMB Stage-IV experiments will generically test thermal dark matter particles with $m \lesssim 15\,\text{MeV}$. We make publicly available the codes developed for this study at \url{https://github.com/MiguelEA/nudec_BSM}.
}

\maketitle

\renewcommand{\baselinestretch}{0.75}\normalsize
\renewcommand{\baselinestretch}{1.0}\normalsize

\section{Introduction}
The number of effective relativistic neutrino species in the early Universe ($N_{\rm eff}$) constitutes a fundamental probe of the thermal history of the early Universe, and as such of the Standard Model of Particle Physics (SM). The agreement between the observations of the Cosmic Microwave Background (CMB)~\cite{Akrami:2018vks,Aghanim:2018eyx}, and the primordial element abundances~\cite{pdg,Cyburt:2015mya} with the SM prediction~\cite{deSalas:2016ztq,Mangano:2005cc,Dolgov:2002wy,Dicus:1982bz,Hannestad:1995rs,Dodelson:1992km,Dolgov:1997mb,Esposito:2000hi,Mangano:2001iu,Birrell:2014uka,Grohs:2015tfy} of three relativistic neutrino species represents a success of the Standard Model, and constitues a non-trivial test for some of its extensions in which light or massless species are present~\cite{Pospelov:2010hj,Iocco:2008va,Sarkar:1995dd,Blennow:2012de,Boehm:2012gr,Boehm:2013jpa,Nollett:2014lwa,Wilkinson:2016gsy,Brust:2013xpv,Vogel:2013raa,Buen-Abad:2015ova,Chacko:2015noa}. In addition, CMB Stage-IV experiments~\cite{Abazajian:2016yjj} are expected to make this test considerably more stringent by achieving sensitivities to $N_{\rm eff}$ of $0.03$, and therefore improving by a factor of $5-6$ current measurements by the Planck satellite~\cite{Akrami:2018vks,Aghanim:2018eyx}.

Given the usefulness of the number of effective neutrino species as a tool to test the SM and its extensions, the neutrino decoupling process in the early Universe has been a subject of intense study~\cite{deSalas:2016ztq,Mangano:2005cc,Dolgov:2002wy,Dicus:1982bz,Hannestad:1995rs,Dodelson:1992km,Dolgov:1997mb,Esposito:2000hi,Mangano:2001iu,Birrell:2014uka,Grohs:2015tfy}. The state of the art modeling of neutrino decoupling~\cite{deSalas:2016ztq,Mangano:2005cc} includes all the relevant physics: exact collision terms, neutrino oscillations, and finite temperature QED corrections, which results in the SM prediction of $N_{\rm eff}^{\rm SM} = 3.045$~\cite{deSalas:2016ztq}. However, although the state of the art treatment of neutrino decoupling in the SM is very accurate, it is considerably involved and computationally expensive since it makes use of the density matrix formalism and requires one to solve a very complex system of kinetic equations\footnote{The system consists on 200 \textit{stiff} coupled integro-differential equations as a result of the binning (in momentum space) of the neutrino distribution functions~\cite{deSalas:2016ztq,Mangano:2005cc,Dolgov:1997mb}.}. On the other hand, studies that have focused on possible BSM implications for $N_{\rm eff}$ with species at the MeV scale have assumed that neutrinos decouple instantaneously and modeled the temperature evolution by assuming entropy conservation (see \textit{e.g.}~\cite{Kolb:1986nf,Serpico:2004nm,Ho:2012ug,Boehm:2012gr,Boehm:2013jpa,Nollett:2014lwa,Wilkinson:2016gsy,Kamada:2015era,Depta:2019lbe}). To the author's knowledge, the sole exception to this approach is~\cite{Kamada:2018zxi}, where a BSM entropy transfer rate between neutrinos and electrons was considered but where the SM electron-neutrino interactions were not taken into account. 

In this study, we carry out a first-principles calculation of $N_{\rm eff}$, in and beyond the Standard Model. We consider all relevant $\nu$-$e$ and $\nu$-$\nu$ interactions, and by assuming thermal equilibrium distribution functions for all the species, negligible chemical potentials, and Maxwell-Boltzmann statistics in the collision terms, we are able to model the neutrino decoupling process in terms of two simple coupled differential equations for $T_\gamma$ and $T_\nu$. We show that this approach has an accuracy of $0.01$ for $N_{\rm eff}$ in the SM, and what is more important: that it allows for a straightforward implementation of BSM states without the need of any a priori knowledge of the time at which neutrinos decouple (which is generically altered by BSM interactions). Therefore, this approach represents a simple -- and yet accurate -- treatment of neutrino decoupling in BSM theories. These two characteristics, accuracy and the ease of BSM species implementation, are crucial in order to make precise theoretical predictions for $N_{\rm eff}$ and the primordial element abundances as synthesized during Big Bang Nucleosynthesis (BBN). In particular, given the upcoming sensitivity of CMB Stage-IV experiments~\cite{Abazajian:2016yjj}.

In this work, we illustrate the usefulness of our simplified approach to neutrino decoupling by revisiting the effect of MeV thermal dark matter (\textit{i.e.} WIMPs~\cite{Bertone:2004pz,Bergstrom:2012fi,Jungman:1995df}) on $N_{\rm eff}$ in a model independent manner. By comparing the theoretical predictions for $N_{\rm eff}$ from our approach with the measured values by the Planck satellite~\cite{Akrami:2018vks,Aghanim:2018eyx} we set bounds on the masses of thermal dark matter particles with $s$-wave and $p$-wave annihilations to electrons and neutrinos. We find that current CMB observations rule out purely electrophilic and neutrinophilic thermal relics with masses $m< 3.0\,\text{MeV}$ at 95\% CL. In addition, we point out that thermal dark matter particles that have non-negligible interactions with both electrons and neutrinos are generically more elusive to CMB observations.

This paper is organized as follows. First, in Section~\ref{sec:NeutrinoDecoupling_BSM} we provide a detailed derivation of the temperature evolution equations that model the relic neutrino decoupling process in the early Universe. This section includes justifications for the various approximations that allow us to capture the physics of the neutrino decoupling process in terms of just two simple differential equations. In Section~\ref{sec:N_eff_DATA}, we review the current and upcoming constraints from CMB observations on $N_{\rm eff}$. In Section~\ref{sec:BSM_neutrinodec}, we work out the time evolution of neutrino decoupling in the presence of MeV scale thermal dark matter particles and set constraints on their masses from CMB observations on $N_{\rm eff}$. In Section~\ref{sec:BBN}, we comment on the complementary bounds that could be inferred from BBN on the dark matter mass. We conclude in Section~\ref{sec:summary}.

\section{Neutrino Decoupling in the Early Universe}\label{sec:NeutrinoDecoupling_BSM}

\subsection{Simplifying approximations for Neutrino Decoupling}\label{sec:assumptions}
 Our aim in this study is not to reproduce the very precise value of $N_{\rm eff}^{\rm SM} = 3.045$ as obtained in Reference~\cite{deSalas:2016ztq}, but to formulate neutrino decoupling in a simplified form that allows one to capture the time dependence of the process while accounting for all possible interactions that can alter it. Which, in addition, thanks to several well justified approximations, allows for a fast, yet reliable and accurate, computation of $N_{\text{eff}}$ -- as relevant for CMB observations -- in the SM and beyond. Our formulation will make use of the integrated Boltzmann equation, and makes the following approximations:
 
 \setlist[itemize]{leftmargin=*,itemsep=1.2pt,leftmargin=0.1cm}
 \begin{enumerate}[leftmargin=0.55cm,itemsep=1.2pt]
 \item
\begin{itemize}
\item[] \textit{Approximation:} the relevant particles follow thermal equilibrium distribution functions \textit{i.e.}: Fermi-Dirac or Bose-Einstein for fermions and bosons respectively.
\item[] \textit{Advantage:} the distribution function of any species is simply determined by two parameters: its temperature $T$ and its chemical potential $\mu$.
\item[] \textit{Justification:} we will be considering particles that at high temperatures are in thermal equilibrium with the SM, and therefore corrections to the equilibrium distributions could only occur when the particles decouple from the plasma. The decoupling happens when the interaction rates are small, and therefore non-thermal spectral distortions are expected to be small. For the case of neutrinos in the SM, the non-thermal corrections amount to $\delta \rho_\nu / \rho_\nu  < 1\,\%$~\cite{deSalas:2016ztq,Mangano:2005cc}. However, we note that if the neutrinos possess any other stronger BSM interactions, the non-thermal corrections may be even smaller. In addition, the electrons, positrons and photons present negligible non-thermal corrections due to the strong electromagnetic interactions.
\end{itemize}

\item 
\begin{itemize}
\item[] \textit{Approximation:} Maxwell-Boltzmann statistics in the collision terms.
\item[] \textit{Advantage:} allows to fully integrate the 6 dimensional collision term for $1 \leftrightarrow 2$ processes, and reduces the 9 dimensional phase space of $2 \leftrightarrow 2$ annihilation processes to just 1.
\item[] \textit{Justification:} the statistical factors in the collision terms modify the rates by less than $15\%$ for thermal distribution functions (see \textit{e.g.}~\cite{Dolgov:1992wf}).
\end{itemize}

 \item 
\begin{itemize}
\item[] \textit{Approximation:} neglect the electron mass in the $e^+ e^- \leftrightarrow \bar{\nu}_i \nu_i$ and $e^\pm \nu_i \leftrightarrow e^\pm \nu_i$ processes.
\item[] \textit{Advantage:} together with Approximations 1 \& 2, permits the full integration of the collision terms, namely, one can write an analytic expression for the energy transfer rates.
\item[] \textit{Justification:} since the electrons are still relativistic when the electron-neutrino interactions freeze-out ($T\sim 2-3\,\text{MeV}$), the error of neglecting the electron mass in the rates is less than a few percent for the relevant temperatures~\cite{Kawasaki:2000en}. 
\end{itemize}

\item
\begin{itemize} 
\item[] \textit{Approximation:} neglect chemical potentials. 
\item[] \textit{Advantage:} reduces the number of evolution equations.
\item[] \textit{Justification:} $\mu_\gamma = \mu_e = 0$ is an excellent assumption due to the very strong scattering and annihilations in the electromagnetic sector. $\mu_{\nu_e} = \mu_{\nu_\mu}  = \mu_{\nu_\tau} = 0$ is also a fairly good approximation since the $e^+ e^- \leftrightarrow \bar{\nu}_i \nu_i$ and $\bar{\nu}_j \nu_j \leftrightarrow \bar{\nu}_i \nu_i$ process damp any possible chemical potentials in the neutrino sector for the temperature range of interest for the process of neutrino decoupling. In addition, we will only consider scenarios in which no primordial leptonic asymmetry exists in the early Universe, \textit{i.e.} $\mu_\nu = \mu_{\bar{\nu}} = 0$.
\end{itemize}

\item
\begin{itemize} \label{assumption:oscillations}
\item[] \textit{Approximation:} the time evolution of neutrino decoupling will be characterized by just two variables $T_\nu = T_{\nu_e} = T_{\nu_\mu} = T_{\nu_\tau}$ and $T_\gamma = T_e$.
\item[] \textit{Advantage:} makes the time evolution of neutrino decoupling extremely simple. 
\item[] \textit{Justification:} $T_\gamma = T_e$ is a very good approximation due to the strong scattering and annihilations between electrons, positrons and photons. $T_{\nu_\mu} = T_{\nu_\tau} $ is also a good assumption because these neutrinos have the same interactions in the SM. Furthermore, given the observed pattern of masses and mixings~\cite{Esteban:2018azc,deSalas:2017kay,Capozzi:2016rtj} the neutrino oscillations become active in the early Universe for temperatures $T= 3-5\,\text{MeV}$~\cite{Hannestad:2001iy,Dolgov:2002ab} (see Appendix~\ref{app:nuoscillations} for more details). The neutrino oscillations act as to equilibrate the flavor distribution functions~\cite{Dolgov:2002ab}. Thus, given the fact that the $\nu_\mu$ and $\nu_\tau$ are -- or are very close to be -- maximally mixed~\cite{Esteban:2018azc,deSalas:2017kay,Capozzi:2016rtj}, $T_{\nu_\mu} = T_{\nu_\tau}$ is also a very good approximation, even in BSM scenarios. In addition, since $\theta_{12},\,\theta_{13},\, \theta_{23} \neq 0$, $T_{\nu_e} \simeq  T_{\nu_{\mu}} \simeq T_{\nu_\tau}$ will be an approximate but almost realistic scenario as a consequence of the neutrino oscillations in the early Universe. Finally, we have explicitly checked that taking $T_{\nu_e}\neq T_{\nu_{\mu,\,\tau}}$ results in very similar values for $N_{\rm eff}$, and although in this work we will assume a common neutrino temperature, the~\href{https://github.com/MiguelEA/nudec_BSM}{online code} allows for a different temperature evolution for each neutrino flavor. 
\end{itemize}
 \end{enumerate}
 
 Collectively, these approximations are well justified and, as we shall see, hugely simplify the time evolution of neutrino decoupling, reducing the system to just two ($dT_\nu/dt, \,dT_\gamma/dt$) ordinary differential equations. In what follows,  we provide the derivation of the time evolution equations for the neutrino and electromagnetic sector temperatures in the early Universe.

\subsection{Modeling of Neutrino Decoupling}
Assisted by the well justified approximations adopted in Section~\ref{sec:assumptions} we can write down two simple evolution equations for $T_\gamma$ and $T_\nu$ that capture the time dependence of the process of neutrino decoupling and that take into account all possible interactions. This will allow for a fast, yet precise, computation of the effective number of relativistic neutrino species as relevant for CMB observations:
\begin{align}\label{eq:N_eff_def_rad}
N_{\rm eff} \equiv  \frac{8}{7} \left(\frac{11}{4}\right)^{4/3} \left(  \frac{\rho_{\rm rad} - \rho_\gamma}{\rho_\gamma} \right) \, .
\end{align} 
Which, when only neutrinos and photons are present, and their distribution functions can be characterized by their temperatures, reads:
\begin{align}\label{eq:N_eff_def}
N_{\rm eff} = 3  \left(\frac{11}{4}\right)^{4/3} \,  \left(\frac{T_{\nu}}{T_\gamma} \right)^4 \, .
\end{align} 
\subsubsection{Derivation of the temperature evolution equations}
The starting point of our derivation is the Liouville equation in a flat, isotropic and homogeneous expanding Universe:
\begin{align}\label{eq:Liouville}
\frac{\partial f}{\partial t}  - H p  \frac{\partial f }{\partial p}  = \mathcal{C}[f]\, ,
\end{align} 
where $f$ is the distribution function of a given species, $p$ is the momentum, $H$ is the Hubble rate:
\begin{align}\label{eq:Hubble}
H  =\sqrt{\frac{8 \pi}{3} \frac{\rho}{m_{Pl}^2}}\,,
\end{align} 
where $m_{Pl} = 1.22\times 10^{19} \, \text{GeV}$, $\rho$ is the total energy density of the Universe, and $\mathcal{C}[f]$ is the collision term that takes into account any potential departure of $f$ from the equilibrium distribution $f^{\rm eq}$ as a result of any scattering/annihilation/decay process. Explicitly, the collision term for a species $\psi$ reads~\cite{Dolgov:2002wy}:
\begin{align}\label{eq:Collisionoperator}
& \mathcal{C}[f_\psi] \equiv - \frac{1}{2 E_\psi} \sum_{X,Y} \int  \prod_i d\Pi_{X_i} \prod_j d\Pi_{Y_j}  (2\pi)^4 \delta^4 (p_\psi + p_X - p_Y)\times  \\
&\left[|\mathcal{M}|^2_{\psi+X\to Y} f_\psi \prod_i f_{X_i} \prod_j \left[1 \pm f_{Y_j} \right] - |\mathcal{M}|^2_{Y\to \psi+X} \prod_j  f_{Y_j}   \left[ 1 \pm f_\psi \right] \prod_i \left[ 1 \pm f_{X_i} \right] \right]\,, \nonumber
\end{align}
where $f_\psi, f_{X_i}, f_{Y_j}$ are the distribution functions of the particles $\psi, \, X_i $ and $Y_j$ respectively. The $-$ sign applies to fermions and the $+$ sign applies to bosons. $d\Pi_{X_i} \equiv \frac{g_{X_i}}{(2 \pi)^3} \frac{d^3 p}{2 E }$ represents the phase space of a particle $X_i$ with $g_{X_i}$ internal degrees of freedom. The delta function ensures energy-momentum conservation, and $\mathcal{M}_{\psi+X\to Y}$ is the amplitude for the $\psi+X\to Y$ process. 

We can integrate Equation~\eqref{eq:Liouville} over momenta with $g\, d^3p/(2\pi)^3$ and $g\,E\,d^3p/(2\pi)^3$ as a measure to obtain:
\begin{align}
 \frac{dn}{dt} + 3 H n &= \frac{\delta n}{\delta t} = \int g  \frac{d^3p}{(2\pi)^3} \, \mathcal{C}[f] \label{eq:n_general} \, , \\
 \frac{d\rho}{dt} + 3 H (\rho + p) &=\frac{\delta \rho}{\delta t} = \int g \,  E \, \frac{d^3p}{(2\pi)^3} \, \mathcal{C}[f] \label{eq:rho_general}  \, ,
\end{align}
where $n$, $\rho$ and $p$ are the number, energy, and pressure densities of the given species respectively. $g$ represents the number internal of degrees of freedom. $\delta n/\delta t$ represents the number density transfer rate, and $\delta \rho/\delta t$ is the energy density transfer rate. Since we neglect chemical potentials, we can write down the temperature evolution of any species just in terms of the energy density evolution~\eqref{eq:rho_general}. Assuming $T_\nu = T_{\nu_e} = T_{\nu_\mu} = T_{\nu_\tau}$, and by using the chain rule, we can write down the temperature evolution equation for the neutrino fluid:
\begin{align}\label{eq:dTnudt}
\frac{dT_\nu}{dt} &= \frac{ -12  H \rho_\nu  + \frac{\delta \rho_{\nu_e}}{\delta t} + 2 \frac{\delta \rho_{\nu_\mu}}{\delta t}}{3 \, \frac{\partial \rho_\nu}{\partial T_\nu }} = - H \, T_\nu + \frac{ \frac{\delta \rho_{\nu_e}}{\delta t} + 2 \frac{\delta \rho_{\nu_\mu}}{\delta t}}{3 \, \frac{\partial \rho_\nu}{\partial T_\nu } }  \, ,
\end{align}
where the first term in the r.h.s. represents the effect of the Universe's expansion, and the second term takes into account non-adiabatic contributions due to the energy exchange between neutrinos and other particles (in the SM, at $T< 10\,\text{MeV}$, electrons and positrons). If $T_{\nu_e} \neq T_{\nu_\mu}$, the temperature evolution equations read:
\begin{subequations}\label{eq:dTnudt_flav}
\begin{align}
\frac{dT_{\nu_e}}{dt} &= \frac{ -4  H \rho_{\nu_e}  + \frac{\delta \rho_{\nu_e}}{\delta t}}{\frac{\partial \rho_{\nu_e}}{\partial T_{\nu_e} }} = - H \, T_{\nu_e} + \frac{ \frac{\delta \rho_{\nu_e}}{\delta t} }{\frac{\partial \rho_{\nu_e}}{\partial T_{\nu_e} } }  \, ,\\
\frac{dT_{\nu_{\mu}}}{dt} &= \frac{ -8  H \rho_{\nu_\mu}  + 2 \frac{\delta \rho_{\nu_\mu}}{\delta t}}{2\frac{\partial \rho_{\nu_\mu}}{\partial T_{\nu_\mu} }} = - H \, T_{\nu_\mu} + \frac{ \frac{\delta \rho_{\nu_\mu}}{\delta t} }{\frac{\partial \rho_{\nu_\mu}}{\partial T_{\nu_\mu} } }  \, .
\end{align}
\end{subequations}
We can obtain the evolution equation for the photon temperature from the continuity equation that relates the net change in the total energy density to the Universe's expansion, the total energy density, and the total pressure~\cite{Kolb:1990vq}:
\begin{align}\label{eq:continuity}
 \dot{\rho}_{\rm Tot}  =- 3 H (\rho_{\rm Tot} + p_{\rm Tot}) \,.
\end{align}
We note that Equation~\eqref{eq:continuity} can be derived both from the fluid perspective ($\nabla_\mu T^{\mu} {}_{\nu} =0  $)~\cite{Kolb:1990vq} and from the microphysics perspective by summing Equation~\eqref{eq:rho_general} over all the species present in the Universe, since $\sum_i  \int g_i\, E_i \, d^3p_i \, \mathcal{C}[f_i] \equiv 0$ (but note that in general $\sum_i \int g_i \, d^3p_i \, \mathcal{C}[f_i] \neq 0 $). 

Finally, from Equations~\eqref{eq:continuity} and~\eqref{eq:rho_general} for the neutrinos we can, by using the chain rule, write down the temperature evolution for the electromagnetic plasma:
\begin{align}\label{eq:T_gamma}
\frac{dT_{\gamma}}{dt}  =- \frac{  4 H \rho_{\gamma} + 3 H \left( \rho_{e} + p_{e}\right) +  \frac{\delta \rho_{\nu_e}}{\delta t} + 2 \frac{\delta \rho_{\nu_\mu}}{\delta t} }{ \frac{\partial \rho_{\gamma}}{\partial T_\gamma} + \frac{\partial \rho_e}{\partial T_\gamma}} \,.
\end{align}
Where again, the first terms in the r.h.s. ($4 H \rho_{\gamma} + 3 H \left( \rho_{e} + p_{e}\right)$) take into account the adiabatic expansion together with the entropy release from electron-positron annihilation while the rest of the terms take into account the non-adiabatic contribution from the energy exchange with other particles (in the SM with neutrinos). This is simply a result of energy conservation: in the SM, the energy exchange rate of the electron-positron fluid is: $\delta\rho_e/\delta t = - \delta\rho_{\nu_e}/\delta t  -2 \, \delta\rho_{\nu_\mu}/\delta t $.

Therefore, by assuming negligible chemical potentials and thermal distributions for the neutrinos, electrons and photons, we are able to model the time evolution of neutrino decoupling in terms of just two coupled differential Equations~\eqref{eq:dTnudt} and~\eqref{eq:T_gamma} subject to the constraint~\eqref{eq:Hubble}.

To finally pose the system we are just left with calculating $\delta \rho_{\nu_e}/\delta t$ and $\delta \rho_{\nu_\mu} /\delta t$ from Equation~\eqref{eq:rho_general}. The processes that contribute to them, in the SM, are $e^+ e^- \leftrightarrow \bar{\nu}_i \nu_i $, $e^\pm \nu_i \leftrightarrow e^\pm \nu_i$, $e^\pm \bar{\nu}_i \leftrightarrow e^\pm \bar{\nu}_i$, $\nu_i \nu_j \leftrightarrow \nu_i \nu_j $,  $\nu_i \bar{\nu}_j \leftrightarrow \nu_i \bar{\nu}_j $, and $\bar{\nu}_i \nu_i \leftrightarrow \bar{\nu}_j {\nu}_j $. The exact collision terms have been calculated in the literature~\cite{Hannestad:1995rs}. Here we use the matrix elements outlined in Table 2 of~\cite{Dolgov:2002wy} and follow the integration method developed in~\cite{Hannestad:1995rs} to obtain the relevant rates\footnote{The reader is referred to the appendices of~\cite{Fradette:2018hhl} and~\cite{Kreisch:2019yzn} which very recently appeared in the literature and which explain in more detail and with explicit formulae how to perform the different integration steps.}. We neglect statistical factors, the electron mass, and assume Maxwell-Boltzmann statistics for all the particles involved in each process to obtain analytic formulae for the rates\footnote{In previous versions of this paper we took the rates from Ref.~\cite{Kawasaki:2000en}. However, we found that the formulae in Ref.~\cite{Kawasaki:2000en} have missing factors of 2 and 8 for annihilation and scattering rates respectively. Such errors have also been reported in footnote 4 of Ref.~\cite{Ichikawa:2005vw}.}. Taking into account all the processes (see Appendix~\ref{app:collision_2body}), the energy exchange rates read:
\begin{subequations}\label{eq:energyrates_nu_SM}
\begin{align}
\left. \frac{\delta \rho_{\nu_e}}{\delta t} \right|_{\rm SM}^{\rm MB} &= \frac{G_F^2}{\pi^5}\left[\left( 1 + 4 s_W^2 + 8 s_W^4 \right) \, F_{\rm MB}(T_\gamma,T_{\nu_e}) + 2 \, F_{\rm MB}(T_{\nu_\mu},T_{\nu_e}) \right] \, ,\\
\left. \frac{\delta \rho_{\nu_\mu}}{\delta t} \right|_{\rm SM}^{\rm MB} &= \frac{G_F^2}{\pi^5}\left[\left( 1 - 4 s_W^2 + 8 s_W^4 \right) \, F_{\rm MB}(T_\gamma,T_{\nu_\mu}) -   F_{\rm MB}(T_{\nu_\mu},T_{\nu_e}) \right] \, ,
\end{align}
\end{subequations}
where we have defined:
\begin{align}
F_{\rm MB}(T_1,T_2) = 32 \, (T_1^9-T_2^9) + 56 \, T_1^4\,T_2^4 \, (T_1-T_2)\, .
\end{align}
and where $G_F= 1.1664\times10^{-5}\,\text{GeV}^{-2}$ is the Fermi constant, and $s_W$ is sinus of the Weinberg angle, which given the observed values of $M_Z$ and $M_W$ reads $s_W^2 = 0.223$~\cite{pdg}.

\subsubsection{Finite temperature QED corrections}

We include the leading order QED finite temperature corrections for the electron and photon mass that result in a smaller pressure of the electromagnetic plasma and hence in an increase of $\sim 0.01$ on $N_{\rm eff}$~\cite{Heckler:1994tv}. We use the finite temperature corrections at leading order in $\alpha$ from~\cite{Heckler:1994tv} and~\cite{Fornengo:1997wa}, and proceed as~\cite{Mangano:2001iu} to obtain the temperature evolution of the electromagnetic plasma including finite temperature corrections:
\begin{align}\label{eq:T_gamma_QED}
\frac{dT_{\gamma}}{dt}  =- \frac{  4 H \rho_{\gamma} + 3 H \left( \rho_{e} + p_{e}\right) + 3 H \, T_\gamma \frac{dP_\text{int}}{dT_\gamma}+ \frac{\delta \rho_{\nu_e}}{\delta t} + 2 \frac{\delta \rho_{\nu_\mu}}{\delta t} }{ \frac{\partial \rho_{\gamma}}{\partial T_\gamma} + \frac{\partial \rho_e}{\partial T_\gamma} +T_\gamma \frac{d^2 P_\text{int}}{dT_\gamma^2} } \,,
\end{align}
where $P_\text{int}$ reads~\cite{Mangano:2001iu}:
\begin{align}\label{eq:P_int}
P_\text{int} = - \int_0^\infty \frac{dp}{2\pi^2} \left[\frac{p^2}{\sqrt{p^2+m_e^2}} \frac{\delta m_e^2(T)}{e^{\sqrt{p^2+m_e^2}/T}+1}+\frac{p}{2} \frac{\delta m_\gamma^2(T)}{e^{p/T}-1} \right] \, ,
\end{align}
and the leading order QED finite temperature correction to the masses are~\cite{Heckler:1994tv,Fornengo:1997wa}:
\begin{align}
\delta m_\gamma^2(T) &=\frac{8\alpha}{\pi} \int_{m_e}^\infty dE  \frac{\sqrt{E^2-m_e^2}}{e^{E/T}+1}\, , \qquad \delta m_e^2(T) = \frac{2\pi \alpha T^2}{3} + \frac{4\alpha}{\pi} \int_{m_e}^\infty dE  \frac{\sqrt{E^2-m_e^2}}{e^{E/T}+1}\,,
\end{align}
where $\alpha$ is the fine structure constant, and for $\delta m_e^2(T)$ we have neglected a momentum dependent contribution that amounts to less than 10\% for an average electron~\cite{Fornengo:1997wa,Mangano:2001iu}. In addition, $\rho_{\rm int} = -P_{\rm int} + T_\gamma \frac{dP_{\rm int}}{dT_\gamma} $ must be included in the Hubble rate~\eqref{eq:Hubble}. 

Note that in practice, since $P_{\rm int}$ is defined as a double integral which evaluation could be slow we simply tabulate it and its derivatives.
\subsubsection{Estimate of the neutrino decoupling temperature in the Standard Model}\label{sec:Tnudec_SM}
The neutrino-electron interactions have a very characteristic energy dependence $\sigma \sim G_F^2 E_\nu^2$, which means that the most energetic neutrinos will decouple later from the electromagnetic plasma. Namely, the neutrino decoupling is not an instantaneous process. However, from Equation~\eqref{eq:dTnudt} we can easily find an estimate for the temperature at which the average neutrino decouples from the electrons and positrons. By expanding $\delta \rho_\nu/\delta t$ in a Taylor series around $\Delta T = T_\gamma -T_\nu$ we find that:
\begin{align}
\frac{dT_\nu}{dt} &   \simeq - H  \left[ T_\nu - \left.  \frac{ \frac{d}{dT_\gamma} \left[\frac{\delta \rho_{\nu_e}}{\delta t} \right] + 2 \frac{d}{dT_\gamma} \left[\frac{\delta \rho_{\nu_\mu}}{\delta t} \right]}{  3 H  \frac{\partial \rho_\nu}{ \partial T_\nu}} \right|_{T_\nu = T_\gamma } \Delta T \right] \,.
\end{align}
And therefore, we notice that the neutrinos will decouple from the plasma, \textit{i.e.} will track the adiabatic expansion of the Universe, when the term accompanying $\Delta T$ becomes $\mathcal{O}(1)$:
\begin{align}\label{eq:deltarho_dt_dec}
\left. \frac{d}{dT_\gamma}\left[\frac{\delta \rho_{\nu}}{\delta t} \right] \right|_{T_\gamma = T_\nu} &\simeq  H  \frac{\partial \rho_\nu}{ \partial T_\nu}\, .
\end{align}
By considering the expressions in~\eqref{eq:energyrates_nu_SM}, $\partial \rho_\nu /\partial T = 4 \rho_\nu/T$, and that only electrons, positrons, photons and neutrinos are present in the early Universe we obtain:
\begin{align}\label{eq:Tnu_dec_SM_vals}
T_{\nu_e}^\text{dec} &\simeq 1.43\,\text{MeV} \, , \qquad \qquad T_{\nu_\mu}^\text{dec} \simeq 2.37\,\text{MeV} \, ,\qquad \qquad T_{\nu}^\text{dec} \simeq 1.83\,\text{MeV} \, .
\end{align}
By numerically integrating the relevant collision terms we find that the neutrino decoupling temperature, including the Fermi-Dirac suppression of the rate, reads:
\begin{align}\label{eq:Tnu_dec_SM_vals_Fermi}
T_{\nu_e}^\text{dec} &\simeq 1.50\,\text{MeV} \, , \qquad \qquad T_{\nu_\mu}^\text{dec} \simeq 2.48\,\text{MeV} \, ,\qquad \qquad T_{\nu}^\text{dec} \simeq 1.91\,\text{MeV} \, .
\end{align}
And hence the decoupling temperatures are enhanced by less than $5\%$ with respect to the MB approximation when the statistical factors are considered.

\subsubsection{$N_{\rm eff}$ in the Standard Model}\label{sec:comparison}

By including the scattering and annihilation interactions between neutrinos and electrons -- together with several well justified approximations -- we have been able to model the process of neutrino decoupling in terms of just two Equations~\eqref{eq:T_gamma_QED} and~\eqref{eq:dTnudt} subject to the constraint~\eqref{eq:Hubble}. Table~\ref{tab:Neff_SM} summarizes the obtained values for $N_\text{eff}$ and the final neutrino temperature in the SM under various assumptions. We note that in the case that we take into account all relevant interactions and finite temperature corrections, we obtain $N_\text{eff}^{\rm SM} = 3.053$ which is remarkably close to the most precise determination in the SM $N_\text{eff}^{\rm SM} = 3.045$~\cite{deSalas:2016ztq}. We attribute the difference between our result and that of Ref.~\cite{deSalas:2016ztq}, to be mainly due to the fact that we do not account for spin-statistics and the electron mass in the interaction rates. Although our modeling does not reproduce the very precise value of $N_\text{eff}^{\rm SM} = 3.045$ obtained in~\cite{deSalas:2016ztq}, a pertinent comment is that our evaluation takes $\lesssim$ 10 s on an average computer, while the evaluation performed in~\cite{deSalas:2016ztq} is computationally very expensive. This fact represents a clear advantage of our approach to explore the impact of BSM scenarios with light particles that can alter the process of neutrino decoupling.

\begin{table}[t]
\begin{center}
\begin{tabular}{cccc}
\hline\hline
Neutrino Decoupling in the SM  	   &$T_\gamma/ T_{\nu}$	& $N_\text{eff}$ \\ \hline
Instantaneous decoupling 		   &	1.4010			&	3	 \\
Instantaneous decoupling + QED &	1.3998			&	3.011	 \\
MB collision term 			    &	1.3961			&  	3.042	 \\
MB collision term + QED  		    &	1.3949			&  	3.053	 \\
  \hline \hline
\end{tabular}
\end{center}
\caption{Summary of the SM results on the number of effective relativistic neutrino species in the early Universe when different effects are taken into account. These numbers should be compared with the most precise evaluation in the SM: $N_\text{eff}^{\rm SM} = 3.045$~\cite{deSalas:2016ztq}. The difference of $\sim 0.01$ results from not accounting for spin-statistics and the electron mass in the interaction rates.}\label{tab:Neff_SM}
\end{table}

\subsection{Numerics}\label{sec:numerics}
In order to solve the system of differential equations for $dT_{\gamma}/dt$~\eqref{eq:T_gamma_QED} and $dT_{\nu}/dt$~\eqref{eq:dTnudt}, in and beyond the SM, we use the \texttt{Odeint} package of \texttt{Python} that is itself based on the \texttt{Fortran} library \texttt{Odepack}. We require the relative and absolute accuracies of the integrator to be $10^{-8}$ and we calculate the pressure and the energy density of massive particles with relative and absolute accuracies of $10^{-8}$. We have explicitly checked that the continuity equation~\eqref{eq:continuity} is fulfilled at each point of the integration with a relative accuracy $< 10^{-5}$ for the SM and all the BSM cases presented in this work. This ensures that \textit{i)} no error has been committed while coding the Hubble rate and the expressions for the temperature evolution, and \textit{ii)} the quoted values of $N_{\rm eff}$ have numerical relative accuracies $< 10^{-5}$.

In all the results presented in this work, we solve the time evolution equations for $T_\gamma$ and $T_\nu$ starting from $T_\gamma = T_\nu = 10\,\text{MeV}$ where the SM weak interactions are highly efficient ($\Gamma/H \sim 50$). The starting time for the evolution is simply $t = 1/(2H)$, which in the SM corresponds to $t_0 \sim 8\times 10^{-3} \, \text{s}$, and we evolve the system until $t_{\rm final} = 5\times 10^4\,\text{s}$. This corresponds to temperatures $T_\gamma \sim 5\times10^{-3}\,\text{MeV}$ at which the electrons and positrons have already annihilated away. 

\section{Current CMB measurements on $N_{\rm eff}$ and upcoming sensitivity}\label{sec:N_eff_DATA}

Both the polarization and temperature CMB spectra are sensitive to the number of effective neutrino species in the early Universe~\cite{pdg,Lesgourgues:2006nd,Lattanzi:2017ubx}. The most obvious effect of a change in $N_{\rm eff}$ is to modify the time of matter-radiation equality which is of pivotal importance for structure formation~\cite{Padmanabhan_struc}. However, such an effect is not unique as can result form a variation in other cosmological parameters ($\Omega_b h^2, \, \Omega_{\rm cdm}h^2$). The most genuine effect of an increase in $N_{\rm eff}$ is to augment the Silk damping scale~\cite{Silk:1967kq}, \textit{i.e.} it amounts to a reduction of the power in the high-$\ell$ spectra~\cite{Bashinsky:2003tk,Hou:2011ec,pdg}. This is how, Stage-IV generation of CMB experiments~\cite{Abazajian:2016yjj} -- that aim to have high resolution at very small angular scales -- expect to reach sensitivities of $\sim 0.03$ on $N_{\rm eff}$. 

In this section we review the current $N_{\rm eff}$ measurements that we should use below to set constraints on MeV scale thermal dark matter particles, and comment on the expected sensitivity from Stage-IV CMB experiments on $N_{\rm eff}$ in order to forecast the reach of Stage-IV CMB experiments to them.

\subsection{Planck 2018}

The Planck satellite~\cite{Akrami:2018vks,Aghanim:2018eyx} reports unprecedented precision measurements of $N_\text{eff}$. Various combinations of data sets have been considered in the Planck 2018 analysis within the $\Lambda$CDM framework, and for our study we shall use the Planck 2018 $N_\text{eff}$ constraints from the combination of TT+TE+EE+lowE+lensing+BAO:
\begin{align}
&N_{\rm eff} = 2.99^{+0.34}_{-0.33}  \qquad \,\,\,\,\, \,\,\,\,\,\,\,\,\,\,\,\,\,\,\, (95\%\,\text{CL, TT+TE+EE+lowE+lensing+BAO}) \, .\label{eq:planck_1} 
\end{align}
In addition, since the observed values of $H_0$ from CMB~\cite{Aghanim:2018eyx} and local observations~\cite{Riess:2016jrr,Riess:2018byc,2018ApJ...855..136R} are in a $\sim 3 \sigma$ tension -- which could potentially be attributed to a BSM $N_{\rm eff}$~\cite{Bernal:2016gxb} -- we also consider the 95\% CL Planck 2018 constraint from the combination of Planck+BAO+$H_0$:
\begin{align}\label{eq:planck_H0}
\,\,N_{\rm eff} &= 3.27 \pm 0.30  \qquad \,\,\,\,\, (95\%\,\text{CL, TT+TE+EE+lowE+lensing+BAO+}H_0) \, .
\end{align}

Given the current disagreement between local and CMB measurements of $H_0$, we consider that an scenario will be conservatively ruled out by Planck 2018 at 95\% CL if $N_{\rm eff} < 2.66$ from TT+TE+EE+lowE or $N_{\rm eff} > 3.57$ from TT+TE+EE+lowE+lensing+BAO+$H_0$.

Finally, we note that generically a change on $N_{\rm eff}$ is also accompanied by a change on the primordial Helium abundance~\cite{Boehm:2013jpa,Nollett:2014lwa,Sarkar:1995dd} ($Y_p$), and the CMB is sensitive to both~\cite{Akrami:2018vks,Aghanim:2018eyx}. However, the combination of Planck observations together with the measurements of the primordial abundances from~\cite{Aver:2015iza} results in $N_{\rm eff} = 2.99^{+0.43}_{-0.40}$ at 95\% CL~\cite{Aghanim:2018eyx}, which is very similar to that from Planck considering only variations of $N_{\rm eff}$~\eqref{eq:planck_1}. Thus, we consider only the Planck constraints in~\eqref{eq:planck_1} but note that they are robust upon modifications of $Y_p$ given the current measurements of the primordial element abundances~\cite{pdg,Cyburt:2015mya,Aver:2015iza}.

\subsection{CMB Stage-IV sensitivity}
CMB Stage-IV experiments aim to measure $N_{\rm eff}$ with a precision of $\sim 0.03$~\cite{Abazajian:2016yjj}. We consider the expected 95\% CL sensitivity by multiplying $0.03$ by $2$. In addition, for consistency, we shall assume as the fiducial measurement of CMB Stage-IV experiments the one we obtain in the SM: $N_{\rm eff} = 3.053$.

\newpage
\section{Neutrino Decoupling BSM: Thermal Dark Matter}\label{sec:BSM_neutrinodec}

A massive stable particle that, in the early Universe, annihilates to light species at a rate of $\left< \sigma v \right>_{\rm WIMP} \simeq 3\times 10^{-26}\,\text{cm}^3/\text{s}$ can account for the observed dark matter abundance in the Universe~\cite{Gondolo:1990dk,Steigman:2012nb}, this is the so-called WIMP miracle~\cite{Kolb:1990vq}. Such a particle would decouple from the plasma while non-relativistic at temperatures of $T \sim m/20$. Therefore, WIMPs of $m\lesssim 20\,\text{MeV}$ will generically alter neutrino decoupling and hence impact $N_{\rm eff}$.

The aim of this section is to work out the temperature evolution equations for the electromagnetic plasma~\eqref{eq:T_gamma_QED} and the neutrino sector~\eqref{eq:dTnudt} in the presence of a WIMP (which will serve as an illustration of how to do so in BSM scenarios in general). From the calculation in the SM (Section~\ref{sec:NeutrinoDecoupling_BSM}) the recipe is clear, if a BSM particle is in thermal equilibrium with either the electromagnetic or neutrino sector of the plasma, then the BSM particle will share its temperature. Generically, BSM particles have interactions with both electrons and neutrinos, so one should include the respective energy exchange rate between the two sectors in Equation~\eqref{eq:rho_general} in order to properly track the temperature evolution.

This Section is structured as follows. First, in subsection~\ref{sec:DM_e_or_nu} we work out the temperature evolution equations when WIMPs, that only annihilate to electrons or neutrinos, are in thermal equilibrium with either electrons or neutrinos respectively. We will compare our results with previous approaches that assumed an instantaneous neutrino decoupling, and we will use CMB constraints on $N_{\rm eff}$ to set a lower bound on the WIMP mass.  In subsection~\ref{sec:DM_e_and_nu} we will consider scenarios in which the WIMP annihilates to both electrons and neutrinos, and will compare the resulting $N_{\rm eff}$ with the observations by the Planck satellite and with the sensitivity of upcoming CMB observations. 

\subsection{Purely electrophilic or neutrinophilic WIMPs}\label{sec:DM_e_or_nu}
A WIMP generically couples to all SM fermions~\cite{Bertone:2004pz,Bergstrom:2012fi,Jungman:1995df}, and therefore, if sufficiently light ($m<m_\mu$), the WIMP annihilation will proceed to neutrinos and electrons. However, for simplicity, in this subsection we will consider scenarios in which WIMPs have couplings to either electrons or neutrinos. The scenario in which WIMPs have couplings to both of these SM fermions will be the case of study of the next subsection. 

\subsubsection{Temperature evolution equations}\label{sec:evol_DM_e_or_nu}
If a WIMP ($\chi$) is solely coupled to the neutrinos, then by summing Equation~\eqref{eq:rho_general} for the neutrinos and the WIMP, and by using the chain rule, the neutrino temperature evolution equation reads:
\begin{align}\label{eq:dTnudt_DM_nu}
\frac{dT_\nu}{dt} &= -\frac{ 12 H \rho_\nu +3 H( \rho_{\chi} + p_{\chi})  - \frac{\delta \rho_{\nu_e}}{\delta t} - 2 \frac{\delta \rho_{\nu_\mu}}{\delta t}}{3 \, \frac{\partial \rho_\nu}{\partial T_\nu } +  \frac{\partial \rho_{\chi}}{\partial T_\nu } } \,,
\end{align} 
while the $T_\gamma$ evolution equation is unaltered and given by Equation~\eqref{eq:T_gamma_QED}. Similarly, for a WIMP that solely interacts with electrons or photons, by summing Equation~\eqref{eq:rho_general} for the WIMP, the photons and the electrons, the photon temperature evolution reads:
\begin{align}\label{eq:T_gamma_DM_e}
\frac{dT_{\gamma}}{dt}  =- \frac{  4 H \rho_{\gamma} + 3 H \left( \rho_{e} + p_{e}\right) +  3 H \left( \rho_{\chi} + p_{\chi}\right) + 3 H \, T_\gamma \frac{dP_\text{int}}{dT_\gamma}+ \frac{\delta \rho_{\nu_e}}{\delta t} + 2 \frac{\delta \rho_{\nu_\mu}}{\delta t} }{ \frac{\partial \rho_{\gamma}}{\partial T_\gamma} + \frac{\partial \rho_e}{\partial T_\gamma} + \frac{\partial \rho_{\chi}}{\partial T_\gamma} +T_\gamma \frac{d^2 P_\text{int}}{dT_\gamma^2} } \,,
\end{align}
while the neutrino temperature evolution is the same as in the SM and given by Equation~\eqref{eq:dTnudt}. Where, of course, it is understood that the energy density of the WIMP contributes to the Hubble expansion~\eqref{eq:Hubble}.

\begin{figure}[t]
\centering
\begin{tabular}{cc}
\hspace{-0.8cm} \includegraphics[width=0.53\textwidth]{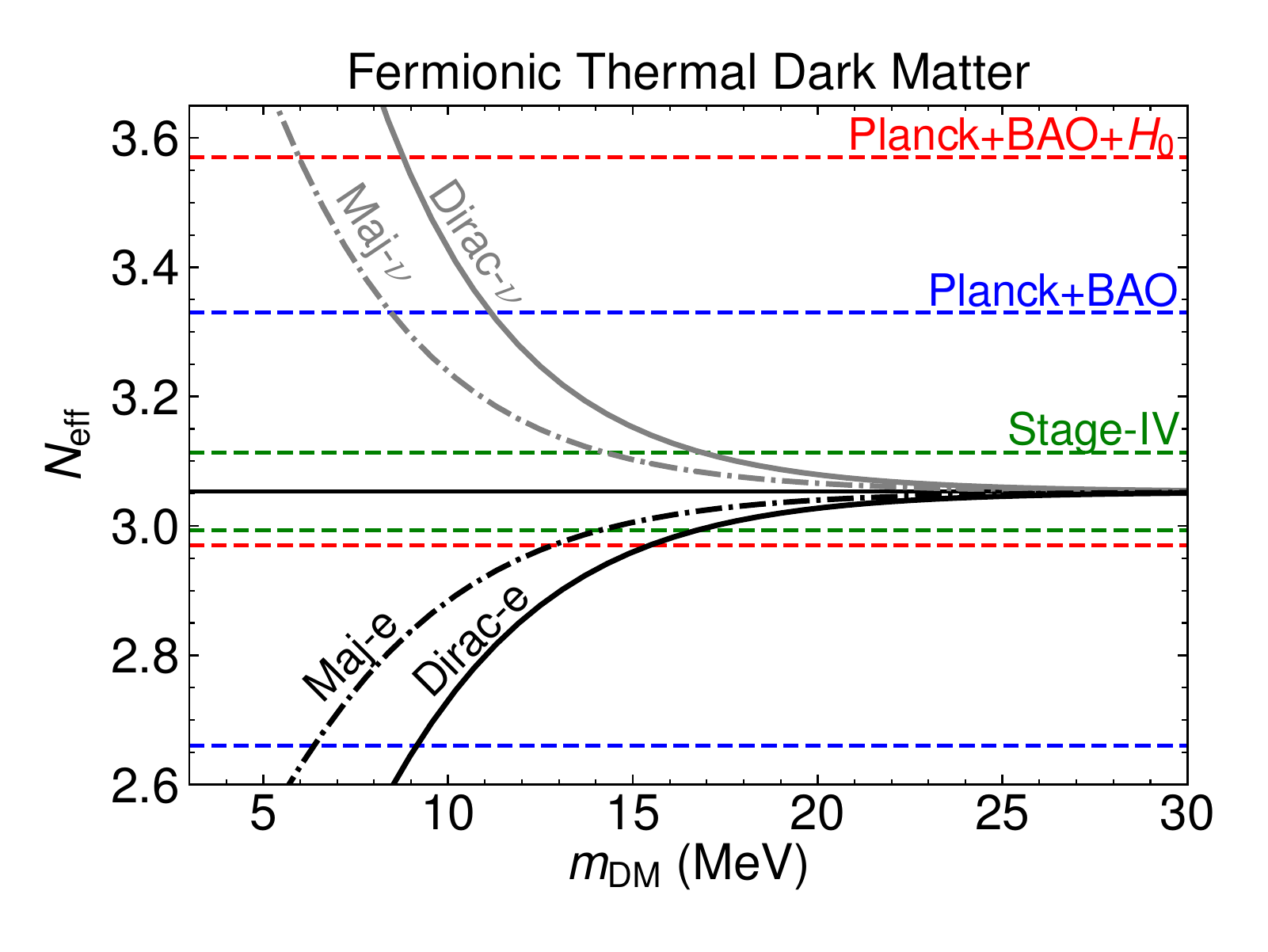} & \hspace{-0.7cm} \includegraphics[width=0.53\textwidth]{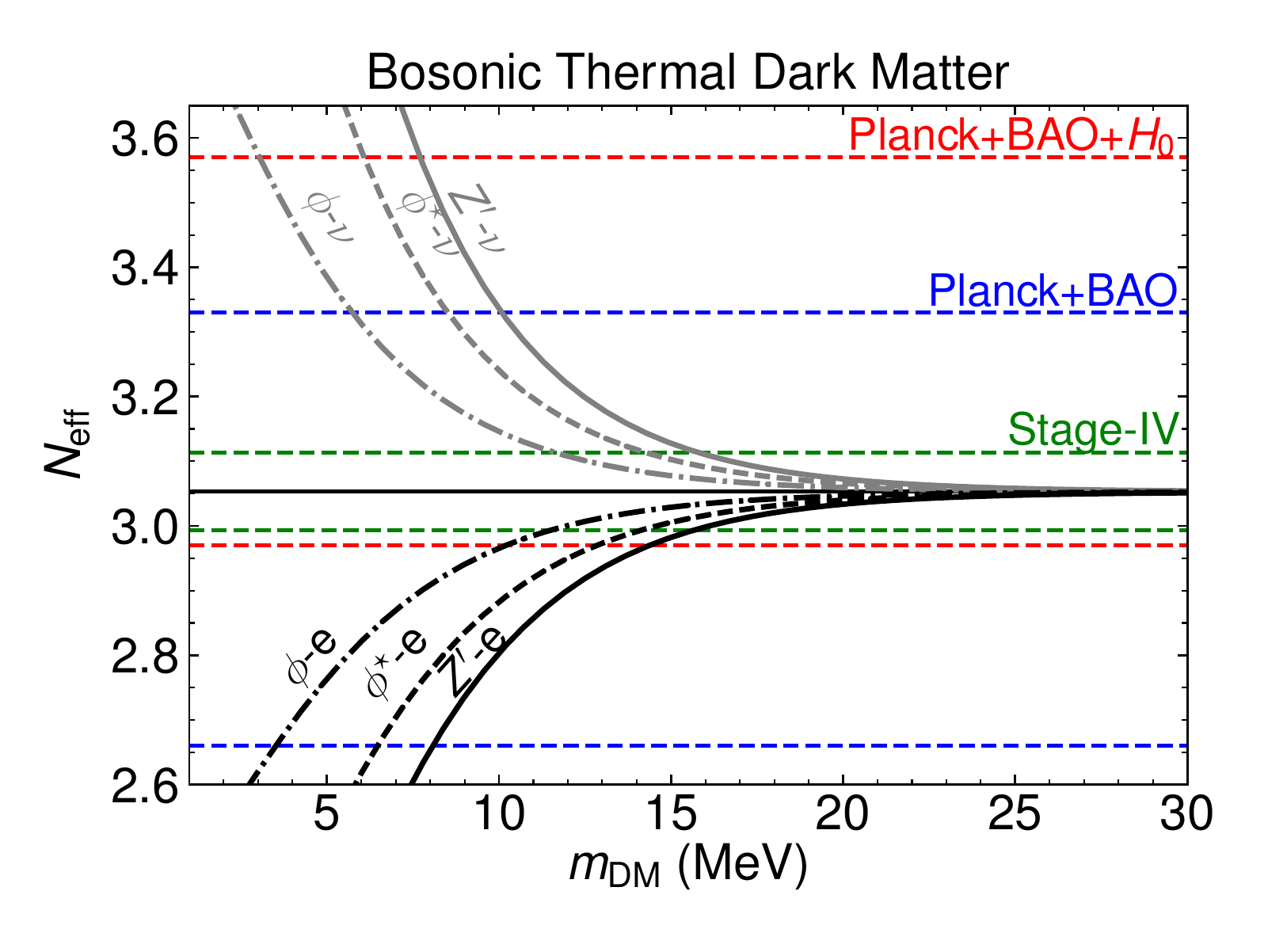}
\end{tabular}
\caption{$N_\text{eff}$ as a function of the mass of a thermal dark matter particle coupled either to electrons or to neutrinos (that apply to both $s$-wave and $p$-wave annihilating relics). \textit{Left panel:} Fermionic dark matter. \textit{Right panel:} Bosonic dark matter: $\phi$, $\phi^\star$ and $Z'$ represent a neutral scalar, charged scalar, and a neutral vector boson respectively. The horizontal dashed lines correspond to the $N_{\rm eff}$ constraints and sensitivity from Planck~\cite{Aghanim:2018eyx} and Stage-IV CMB experiments~\cite{Abazajian:2016yjj} respectively. See Table~\ref{tab:SUMMARY} for the lower bounds on the masses in each case. }\label{fig:DM_Neff}
\end{figure}
\subsubsection{Results and discussion}\label{sec:res_DM_e_or_nu}
The energy density and pressure of a given WIMP are uniquely determined by its internal number of degrees of freedom $g $, its spin, and its temperature. In Figure~\ref{fig:DM_Neff} we show the resulting $N_{\rm eff}$ -- for fermionic and bosonic dark matter particles -- as a function of the dark matter mass for various choices of $g$ and spins. Electrophilic relics release entropy into the electromagnetic plasma and hence result in $N_{\rm eff} < N_{\rm eff}^{\rm SM}$ (see Equation~\eqref{eq:N_eff_def}), while neutrinophilic relics release entropy into the neutrino sector and hence result in $N_{\rm eff} > N_{\rm eff}^{\rm SM}$. In Table~\ref{tab:SUMMARY} we provide a summary of the current constraints and future sensitivities for the masses of WIMPs with various spins, and that couple to either electrons or neutrinos. We note that, irrespectively of the spin of the WIMP, Planck observations rule out electrophilic and neutrinophilic thermal dark matter particles with $m  < 3.0\,\text{MeV}$ at 95\% CL. In addition, we show that thermal relics purely annihilating to electrons or neutrinos with $m  \lesssim 15\,\text{MeV}$ will be generically tested by upcoming CMB experiments. 

We note that the obtained lower bound on the dark matter mass is independent on the low velocity behaviour of the annihilation process. This is because the quantity that controls the WIMP relic abundance is the thermally averaged annihilation cross section, and this should be $\left< \sigma v \right> \simeq 3\times 10^{-26}\,\text{cm}^3/\text{s}$ regardless of the behaviour at low velocities~\cite{Gondolo:1990dk,Steigman:2012nb,Kolb:1990vq}. This makes our bounds particularly relevant for $p$-wave annihilating relics to electrons since CMB observations are not sensitive to them~\cite{Aghanim:2018eyx,Slatyer:2015jla}. This will be for instance the case of a Majorana dark matter particle annihilating to electrons via a dark photon exchange, or a Dirac dark matter particle annihilating to electrons via a dark Higgs exchange~\cite{Kumar:2013iva,Krnjaic:2015mbs,Alexander:2016aln,Battaglieri:2017aum}. Our bounds are of relevance for both $s$-wave and $p$-wave annihilating WIMPs to neutrinos~\cite{Boehm:2006mi,Farzan:2009ji,Farzan:2011ck,Batell:2017cmf,Berlin:2018sjs} that remain elusive to searches at terrestrial neutrino experiments~\cite{Beacom:2006tt,Campo:2017nwh,Klop:2018ltd}.

\begin{table}[t]
\begin{center}
{\def\arraystretch{1.4}
\begin{tabular}{lc|cc|cc|cc}
\hline\hline
   	 \multicolumn{2}{c|}{Dark Matter}   	 &   \multicolumn{6}{c}{$N_{\rm eff}$ Constraints } \\ \cline{3-8} 
   	 \multirow{2}{*}{Particle}    	 &   \multirow{2}{*}{g-Spin}  &   \multicolumn{2}{c|}{Planck+BAO}  &   \multicolumn{2}{c|}{Planck+BAO+$H_0$}  &   \multicolumn{2}{c}{Stage-IV Sensitivity}      \\ \cline{3-8} 

   &  &		  				 	$> 2.66$   &	$ < 3.33$	   	&  $>2.97$	   & $ < 3.57$   &  $ > 2.99$ & $ <3.11$ \\ 
  \hline \hline
Majorana DM-$\nu$  &	2-F		&  - &	 8.5            &   - & 6.0           &   - &14.3          \\\hline
Dirac DM-$\nu$  	&	4-F		&  - &  	  11.2           &  - &  8.8            &  - &   16.9           \\\hline
$\phi$ DM-$\nu$  &		1-B		&  - &  	   5.7          &  - &    3.0          &  - &      11.7        \\\hline
$\phi^\star$ DM-$\nu$  &	2-B		&  - &  	   8.5          &  - &  6.1            &  - &     14.3        \\\hline
$Z'$ DM-$\nu$  &		3-B		&  - &  	  10.1           &  - &    7.7          &  - &    15.8          \\\hline
\hline
Majorana DM-$e$  &		2-F		&  	   6.4          &  - &      12.8         &  - &      14.1     &  -    \\\hline
Dirac DM-$e$  &		4-F		&   	 9.2	            &  - &     15.5          &  - &       16.7  &  -      \\\hline
$\phi$ DM-$e$  &		1-B		&  	3.5             &  - &     10.2          &  - &       11.5  &  -      \\\hline
$\phi^\star$ DM-$e$  &	2-B		&  	 6.5            &  - &    12.9           &  - &      14.1   &  -      \\\hline
$Z'$ DM-$e$  &			3-B		&  	  8.1           &  - &    14.4           &  - &       15.6   &  -     	 \\
  \hline \hline
\end{tabular}
}
\end{center}
\vspace{-0.2cm}
\caption{Lower bounds on the masses of various thermal dark matter particles (in MeV) at 95\% CL.  $\nu$ refers to neutrinophilic dark matter particles, and $e$ refers to electrophilic dark matter particles. Given the current disagreement between local and CMB measurements of the Hubble constant, a conservative bound would be to consider the smallest number from the Planck+BAO and Planck+BAO+$H_0$ columns. Note that the bounds apply to both $s$-wave and $p$-wave annihilating  relics. }\label{tab:SUMMARY}
\end{table}

\newpage
\subsubsection{Comparison with previous literature}\label{sec:compa_DM_e_or_nu}
\begin{figure}[t]
\centering
\begin{tabular}{cc}
\hspace{-0.8cm} \includegraphics[width=0.53\textwidth]{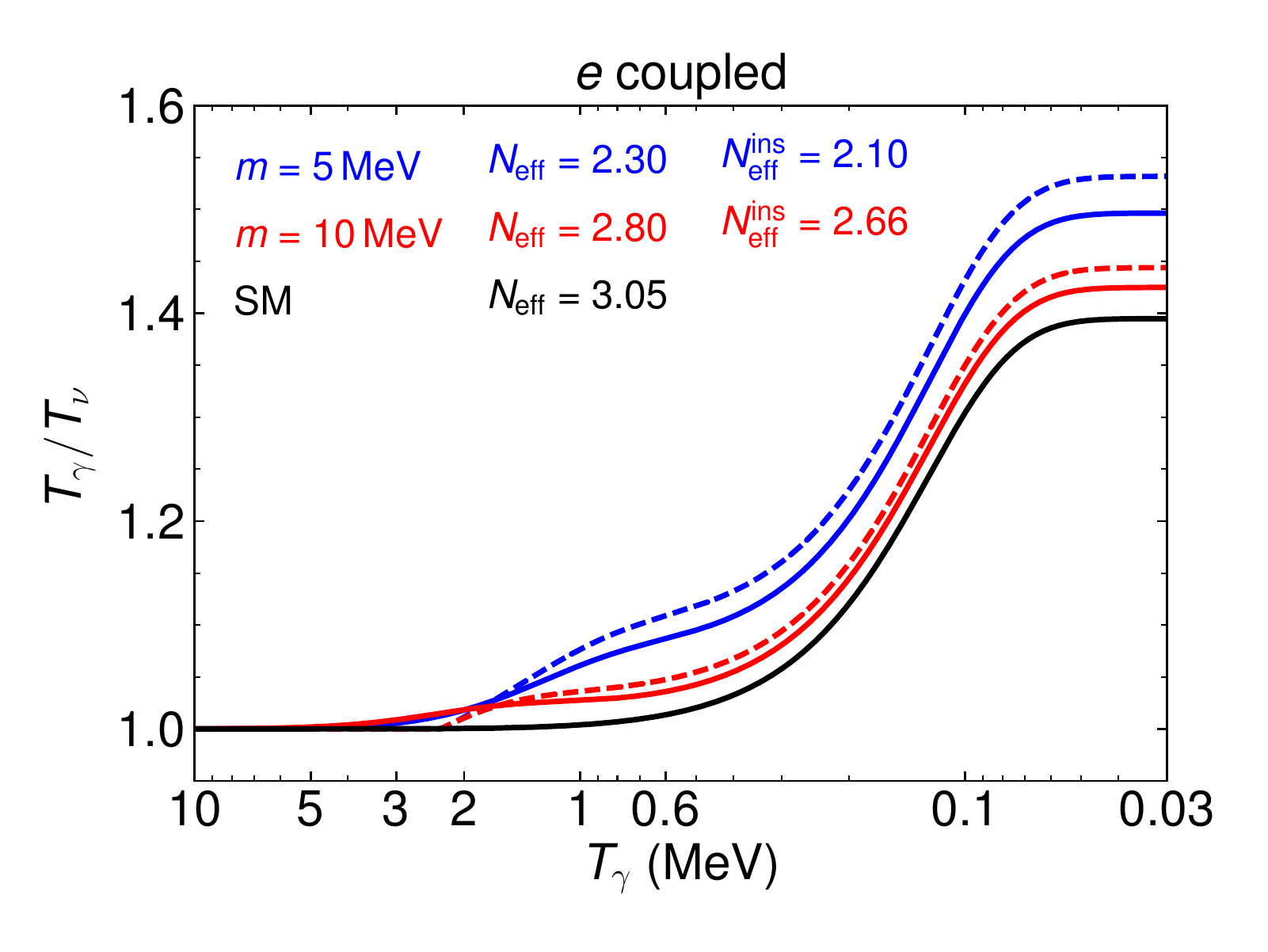} & \hspace{-0.7cm} \includegraphics[width=0.53\textwidth]{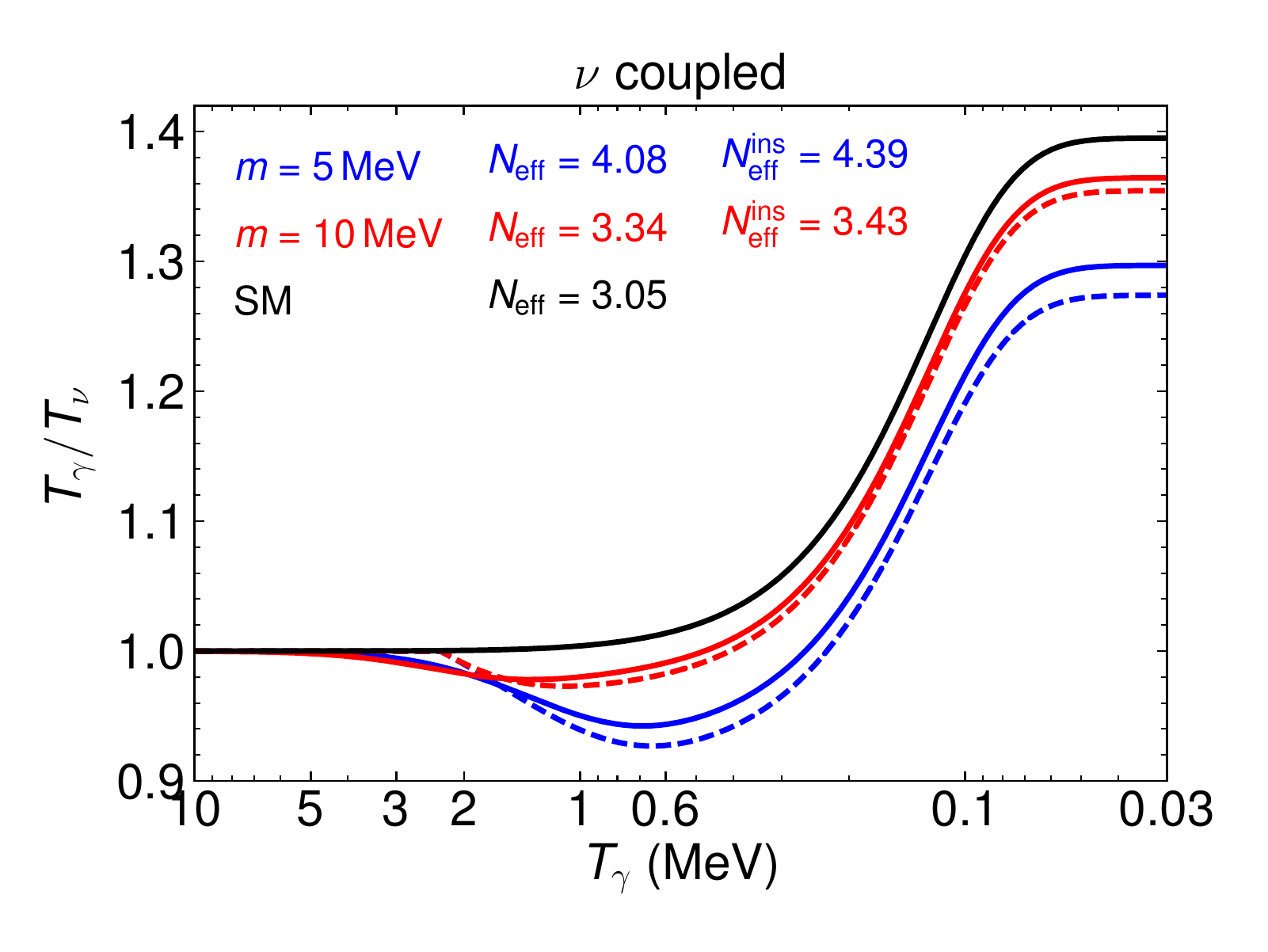} 
\end{tabular}\vspace{-0.1cm}
\caption{Temperature evolution of the neutrino sector as a function of the photon temperature including a neutral vector boson in thermal equilibrium with either the electromagnetic  (\textit{left panel}) or neutrino (\textit{right panel}) component of the plasma. The solid lines result from solving our derived system of equations:~\eqref{eq:T_gamma_DM_e},~\eqref{eq:dTnudt}, and~\eqref{eq:dTnudt_DM_nu},~\eqref{eq:T_gamma_QED} for the left and right panels respectively. The dashed lines correspond to the neutrino temperature evolution as obtained in the instantaneous decoupling approximation with $T_{\nu}^\text{dec} = 2.3\,\text{MeV}$ as commonly used in the literature. }\label{fig:comparison}
\end{figure}

The impact on neutrino decoupling of MeV scale particles coupled to either the electromagnetic plasma or the neutrino sector has been studied in the past~\cite{Kolb:1986nf,Serpico:2004nm,Ho:2012ug,Boehm:2012gr,Boehm:2013jpa,Nollett:2014lwa,Wilkinson:2016gsy}. These studies considered the effect of such particles by assuming that neutrinos decouple instantaneously at a temperature of $T_{\nu}^{\rm dec} =2.3\,\text{MeV}$, and then they tracked the temperature of the neutrino and electromagnetic sectors by imposing entropy conservation~\cite{Kolb:1986nf}. 

Figure~\ref{fig:comparison} displays a direct comparison of the neutrino temperature evolution from our approach (solid lines) and the previously used in the literature (dashed lines). We clearly notice that the traditional approach shows a somewhat artificial behaviour of the neutrino temperature evolution in the region $T = 1- 5\,\text{MeV}$ as a result of imposing entropy conservation. Our approach, on the other hand, describes the temperature evolution of the neutrino fluid with a smooth evolution that is attributed to the continuous entropy release of the BSM particle into the sector to which it interacts with. 

We also appreciate a notable difference on $N_{\rm eff}$, which could be as large as $\sim 0.2$. At the moment, this difference is below current CMB experiments sensitivities, however, for CMB Stage-IV experiments that aim at sensitivity of $0.03$ for $N_{\rm eff}$, the comparison would result in biased results. The difference on $N_{\rm eff}$ from our approach and the traditional one that assumes an instantaneous decoupling is a result of two effects. First, in the Maxwell-Boltzmann approximation, the neutrinos decouple at $T \simeq 1.83\,\text{MeV}$~\eqref{eq:Tnu_dec_SM_vals} and not at $T=2.3\,\text{MeV}$ (while in reality they do so at $T\simeq 1.91\,\text{MeV}$~\eqref{eq:Tnu_dec_SM_vals_Fermi}). The earlier the neutrinos decouple, the larger the effect on $N_{\rm eff}$ is from a massive relic coupled to either the electromagnetic or neutrino sector. And second, our approach takes into account finite temperature QED corrections that amount to $\Delta N_{\rm eff} = 0.011$. We have explicitly checked that the bounds presented in Table~\ref{tab:SUMMARY} are $7$-$16\,\%$, $14$-$24\,\%$, $\sim$$-5\%$ less stringent than those that would be obtained using the traditional instantaneous decoupling approximation for purely neutrinophilic WIMPs (where each number corresponds to Planck+BAO, Planck+BAO+$H_0$, and Stage-IV respectively). Similarly, for WIMPs coupled to electrons, our bounds are $24$-$50\,\%$, $18$-$36\,\%$, $21$-$38\,\%$ more restrictive than those obtained within the traditional instantaneous decoupling approximation. These modifications can be understood from the difference between the traditionally used neutrino decoupling temperature, $T_{\nu}^{\rm dec} =2.3\,\text{MeV}$, and the actual neutrino decoupling temperature within the MB approximation, $T \simeq 1.83\,\text{MeV}$~\eqref{eq:Tnu_dec_SM_vals} -- which is roughly $30\%$.

Finally, we note that although the differences between our approach and the commonly used in the literature are small (up to 25\%) for $\Delta N_{\rm eff}$, the key possibility that our approach offers for BSM scenarios is that one can easily include energy exchange rates between sectors. Namely, we can consider what is the effect of WIMPs that simultaneously interact with electrons and neutrinos, which is generically what happens~\cite{Bertone:2004pz,Bergstrom:2012fi,Jungman:1995df}. This is the case of study of the next subsection.

\subsection{Thermal Dark Matter annihilating to electrons and neutrinos}\label{sec:DM_e_and_nu}
The couplings of a WIMP with mass $m_\chi \lesssim 30\,\text{MeV}$ to electrons and neutrinos will vary in a model to model basis~\cite{Bertone:2004pz,Bergstrom:2012fi,Jungman:1995df}. In this section, we consider the case of a WIMP with couplings to these two SM fermions in order to illustrate how  can be affected by a generic WIMP. We note that this scenario has not previously been addressed in the literature and serves as a neat example of the possibilities that our simplified approach to the process of neutrino decoupling offers.

\subsubsection{Temperature evolution equations}\label{sec:evol_DM_e_and_nu}
The WIMP ($\chi$) will share the temperature of the sector to which it annihilates the most. So, if the annihilation predominantly proceeds to neutrinos in the early Universe, the temperature evolution equations will read:
\begin{subequations}\label{eq:T_gamma_DM_a}
\begin{align}
\frac{dT_\nu}{dt} &= -\frac{ 12 H \rho_\nu +3 H( \rho_\chi + p_\chi)  - \frac{\delta \rho_{\nu_e}}{\delta t} - 2 \frac{\delta \rho_{\nu_\mu}}{\delta t}  - \left. \frac{\delta \rho_\chi}{\delta t} \right|_{e} }{3 \, \frac{\partial \rho_\nu}{\partial T_\nu } +  \frac{\partial \rho_\chi}{\partial T_\nu } } \,, \\
\frac{dT_{\gamma}}{dt}  &=- \frac{  4 H \rho_{\gamma} + 3 H \left( \rho_{e} + p_{e}\right) + 3 H \, T_\gamma \frac{dP_\text{int}}{dT_\gamma}+ \frac{\delta \rho_{\nu_e}}{\delta t} + 2 \frac{\delta \rho_{\nu_\mu}}{\delta t} + \left. \frac{\delta \rho_\chi}{\delta t} \right|_{e} }{ \frac{\partial \rho_{\gamma}}{\partial T_\gamma} + \frac{\partial \rho_e}{\partial T_\gamma} +T_\gamma \frac{d^2 P_\text{int}}{dT_\gamma^2} } \,.
\end{align}
\end{subequations}
Where these equations differ from~\eqref{eq:dTnudt_DM_nu} and~\eqref{eq:T_gamma_QED} in that we have allowed for a non-vanishing energy transfer rate between the neutrino and electromagnetic components of the plasma, as a result of the dark matter particle interactions: $\delta \rho_\chi/\delta t$. 

In a similar spirit, if the annihilation proceeds predominantly to electrons in the early Universe, the temperature evolution equations reads:
\begin{subequations}\label{eq:T_gamma_DM_b}
\begin{align}\frac{dT_{\gamma}}{dt}  &=- \frac{  4 H \rho_{\gamma} + 3 H \left( \rho_{e} + p_{e}\right) +  3 H \left( \rho_\chi + p_\chi\right) + 3 H \, T_\gamma \frac{dP_\text{int}}{dT_\gamma}+ \frac{\delta \rho_{\nu_e}}{\delta t} + 2 \frac{\delta \rho_{\nu_\mu}}{\delta t} - \left. \frac{\delta \rho_\chi}{\delta t} \right|_{\nu} }{ \frac{\partial \rho_{\gamma}}{\partial T_\gamma} + \frac{\partial \rho_e}{\partial T_\gamma} + \frac{\partial \rho_\chi}{\partial T_\gamma} +T_\gamma \frac{d^2 P_\text{int}}{dT_\gamma^2} } \,, \\
\frac{dT_\nu}{dt} &= -\frac{ 12 H \rho_\nu - \frac{\delta \rho_{\nu_e}}{\delta t} - 2 \frac{\delta \rho_{\nu_\mu}}{\delta t} +  \left. \frac{\delta \rho_\chi}{\delta t} \right|_{\nu} }{3 \, \frac{\partial \rho_\nu}{\partial T_\nu } } \,.
\end{align}
\end{subequations}
To solve these systems of equations, we are just left with calculating the energy exchange rates as induced by the dark matter interactions. Unfortunately, these energy exchange rates are model dependent. This is because $\delta \rho_\chi/{\delta t}$ takes into account both scattering and annihilation interactions, but the condition of obtaining the correct relic abundance only fixes the annihilation rate to be $\left< \sigma v \right> \simeq 3\times 10^{-26}\,\text{cm}^3/\text{s}$~\cite{Gondolo:1990dk,Steigman:2012nb}. The parametric dependence on the couplings and masses of the scattering and annihilation cross sections are similar. However, although the energy exchanged in a scattering process $\sim T$ can be substantially smaller to that in an annihilation $\sim m + 3\,T$, the annihilation rate should be suppressed by a factor $\sim  e^{-m/T}$ with respect to the scattering rate because the light species to which the WIMP scatters are relativistic. Here, we consider only the annihilation energy exchange rate but note that for any given model one should also include the scattering rates (yet, this will serve as a clear illustration of the phenomenology of this scenario). We calculate the energy exchange rate in a model independent manner by considering the usual WIMP number density evolution (this is simply Equation~\eqref{eq:n_general} for a stable particle)~\cite{Kolb:1990vq,Gondolo:1990dk}:
\begin{align}\label{eq:dn_WIMP}
 \frac{dn_{\chi}}{dt} + 3 H n_{\chi} &= -\left< \sigma v \right> \left( n_{\chi}^2 - n^2_{\chi} {}^{\rm eq}\right)\, ,
\end{align}
where $n^{\rm eq}$ corresponds to the equilibrium number density. We can easily calculate the energy exchange rate by noting that the energy exchanged in the annihilation process will simply be $\sim 2 m_{\chi}$ for a non-relativistic relic ($T<m_\chi/3$). Thus, the energy exchange rate is:
\begin{align}\label{eq:drho_WIMP}
 \frac{\delta \rho_\chi}{\delta t} &\simeq m_\chi  \frac{\delta n_\chi}{\delta t} \simeq - m_\chi\left< \sigma v \right> \left( n_\chi^2 - n^2_\chi {}^{\rm eq}\right)\, .
 \end{align}
Which, for WIMPs that are in thermal equilibrium with neutrinos but that also annihilate to electrons ($\left< \sigma v \right>_{ \chi \chi \to e^+ e^-} \leq \left< \sigma v \right>_{ \chi \chi \to \bar{\nu} \nu}$), in the MB approximation reads:
 \begin{align}\label{eq:drho_WIMP_e}
\left.  \frac{\delta \rho_\chi}{\delta t} \right|_e &=  \frac{g_\chi^2 m_{\rm \chi}^5}{4\pi^4} \, \left< \sigma v \right>_{ \chi \chi \to e^+ e^-}  \left[  T_\gamma^2 \, K_2^2\left[\frac{m_{\chi}}{T_{\gamma}}\right] - T_\nu^2 \, K_2^2\left[\frac{m_\chi}{T_{\nu}}\right] \right]  \, ,
\end{align}
where $K_2$ is a modified Bessel function of the second kind. For WIMPs that are in thermal equilibrium with the electromagnetic plasma but that also have interactions with neutrinos, in the MB approximation reads:
 \begin{align}\label{eq:drho_WIMP_nu}
\left.  \frac{\delta \rho_\chi}{\delta t} \right|_\nu &= \frac{g_\chi^2 m_{\rm \chi}^5}{4\pi^4} \, \left< \sigma v \right>_{\chi \chi \to \bar{\nu}\nu} \left[T_\nu^2  \, K_2^2\left[\frac{m_{\chi}}{T_{\nu}} \right] -  T_\gamma^2 \, K_2^2\left[\frac{m_{\chi}}{T_{\gamma}} \right] \right]  \,.
\end{align}
Note that the annihilation cross section for WIMPs that annihilate to electrons and neutrinos is fixed to be $\left< \sigma v \right>_{\rm WIMP} = \left< \sigma v \right>_e + \left< \sigma v \right>_\nu  \simeq 3\times 10^{26}\,\text{cm}^3/\text{s}$.

\begin{figure}[t]
\centering
\begin{tabular}{cc}
\hspace{-0.8cm} \includegraphics[width=0.53\textwidth]{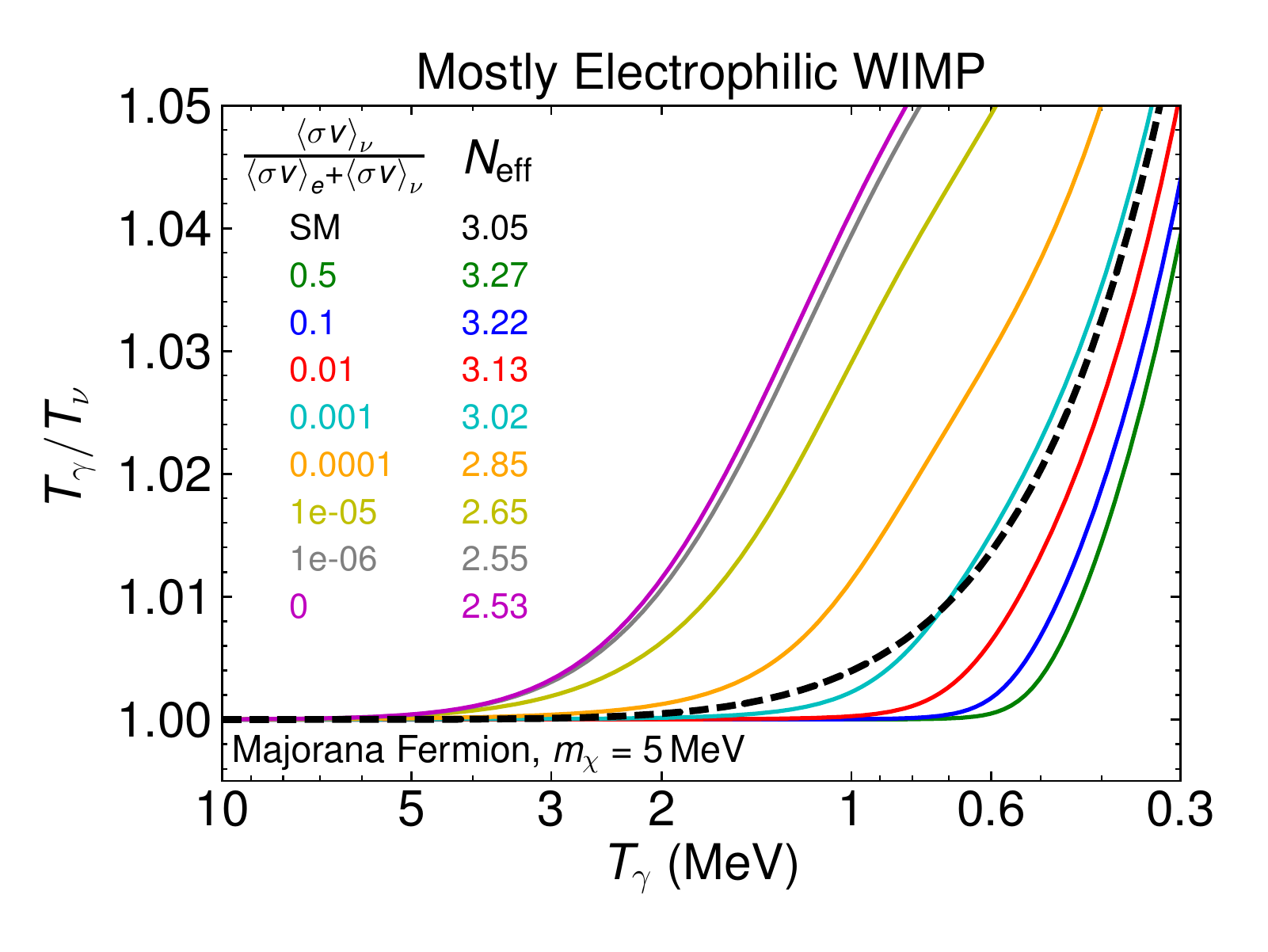} & \hspace{-0.7cm} \includegraphics[width=0.53\textwidth]{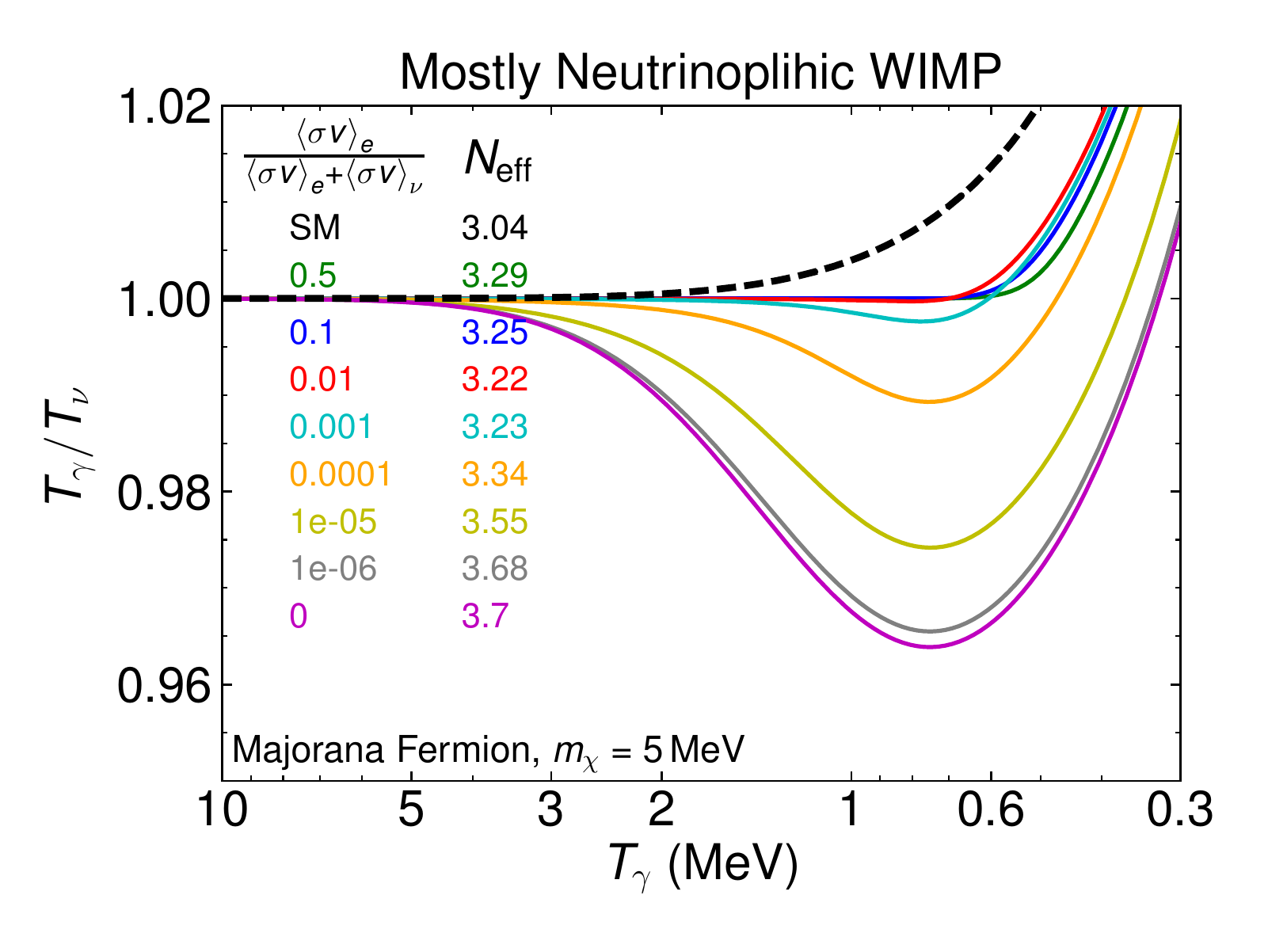} 
\end{tabular}\vspace{-0.3cm}
\caption{Neutrino temperature evolution as a function of the photon temperature when a light Majorana WIMP with $m_\chi = 5\,\text{MeV}$ is in thermal equilibrium in the early Universe. \textit{Left panel:} WIMP that annihilates primarily to electrons. \textit{Right panel:} WIMP that annihilates primarily to neutrinos. Note that apart from releasing entropy from its annihilations, the effect of a WIMP with sizeable annihilations to both electrons and neutrinos is to maintain the electrons and neutrinos coupled for smaller temperatures than in the SM, $T_{\nu}^{\rm dec} \simeq 1.8\,\text{MeV}$. }\label{fig:TemperatureEvolution_WIMP}
\end{figure}

\subsubsection{Results and discussion}\label{sec:evol_DM_e_and_nu}
In Figure~\ref{fig:TemperatureEvolution_WIMP} we show the temperature evolution of the neutrino sector as a function of the photon temperature in the presence of a WIMP that annihilates to neutrinos and electrons. The various lines correspond to different final state ratios of the annihilation process. We notice that the effect of light WIMPs that have sizeable interactions with both electrons and neutrinos ($\left< \sigma v \right>_{e,\, \nu}/\left< \sigma v \right>_{\rm WIMP} \gtrsim 10^{-5}$) is to maintain the neutrinos coupled to the electromagnetic plasma for smaller temperatures than the expected one in the SM, $T_{\nu}^{\rm dec} \simeq 1.8\,\text{MeV}$~\eqref{eq:Tnu_dec_SM_vals}. To have an understanding of why this is the case, we can actually estimate the neutrino decoupling temperature by following a similar approach to that in the SM (see Section~\ref{sec:Tnudec_SM}), namely, we can compare the change in the energy exchange rate from the WIMP annihilation to the Hubble expansion:
 \begin{align}\label{eq:WIMP_rho_dec}
\left. \frac{d}{dT_\gamma}\left[\frac{\delta \rho_{\chi}}{\delta t} \right] \right|_{T_\gamma = T_\nu} &\simeq  H  \frac{\partial \rho_\nu}{ \partial T_\nu}\, .
\end{align}
By considering Equation~\eqref{eq:drho_WIMP_e} and approximating $K_2\left[\frac{m}{T}\right]\simeq \sqrt{\frac{\pi T}{2m}} \, e^{-m/T}$ we can easily find the temperature at which the energy transfer between electrons and neutrinos as a result of the WIMP annihilations becomes inefficient:
\begin{align}\label{eq:dec_invdecay}
\frac{m_\chi}{T} \simeq \frac{1}{2} \log\left[\frac{ 5\, g_\chi^2 \left< \sigma v \right>_{\chi \chi \to \bar{\nu}\nu} m_\chi m_{Pl}}{1.66 \times 14 \, \pi^5\,\sqrt{g_\star}}\right] + 2 \log\left[\frac{m_{\chi}}{T}\right] \, .
\end{align}
Where $g_\star$ corresponds to the number of relativistic effective degrees of freedom contributing to radiation. We notice that the decoupling temperature dependence with the annihilation cross section is only logarithmic. Plugging in representative numbers, we find:
\begin{align}\label{eq:dec_invdecay_num}
\frac{m_\chi}{T} \simeq 6.05 + \frac{1}{2} \log\left[\frac{\left< \sigma v \right>_{\chi \chi \to \bar{\nu}\nu}}{10^{-3} \times \left< \sigma v \right>_{\rm WIMP}} \frac{5\,\text{MeV}}{m_{\chi}}  \frac{g_\chi^2}{4} \sqrt{\frac{10.75}{g_\star}}\right] + 2 \log\left[\frac{m_{\chi}/T}{6.05}\right]  \, .
\end{align}
Thus, we realize that a light WIMP that mainly annihilates to electrons but that only annihilates to neutrinos once every $10^3$ times can still significantly alter the temperature at which neutrinos decouple. For a WIMP with $m_\chi = 5\,\text{MeV}$, $\left< \sigma v \right>_e = (1-10^{-3})\left< \sigma v \right>_{\rm WIMP}$, and $\left< \sigma v \right>_\nu = 10^{-3}\left< \sigma v \right>_{\rm WIMP}$, then $T_{\nu}^{\rm dec} \sim 0.8\,\text{MeV}$, which is substantially smaller than the temperature at which neutrinos decouple in the SM $T_{\nu}^{\rm dec} \sim 1.8\,\text{MeV}$~\eqref{eq:Tnu_dec_SM_vals}. Thus, we notice that the effect of WIMPs with masses $m\lesssim 20\,\text{MeV}$ is not only to release entropy into the system but they generically act as to delay the time of neutrino decoupling.

\begin{figure}[t]
\centering
\hspace{1.4cm} \includegraphics[width=0.7\textwidth]{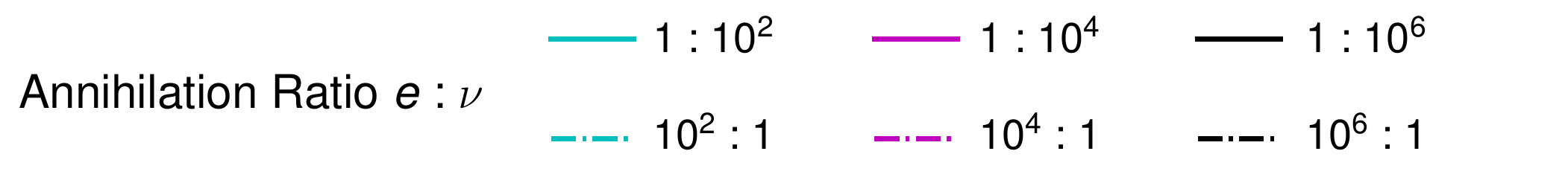}
\begin{tabular}{cc}
\hspace{-0.8cm} \includegraphics[width=0.53\textwidth]{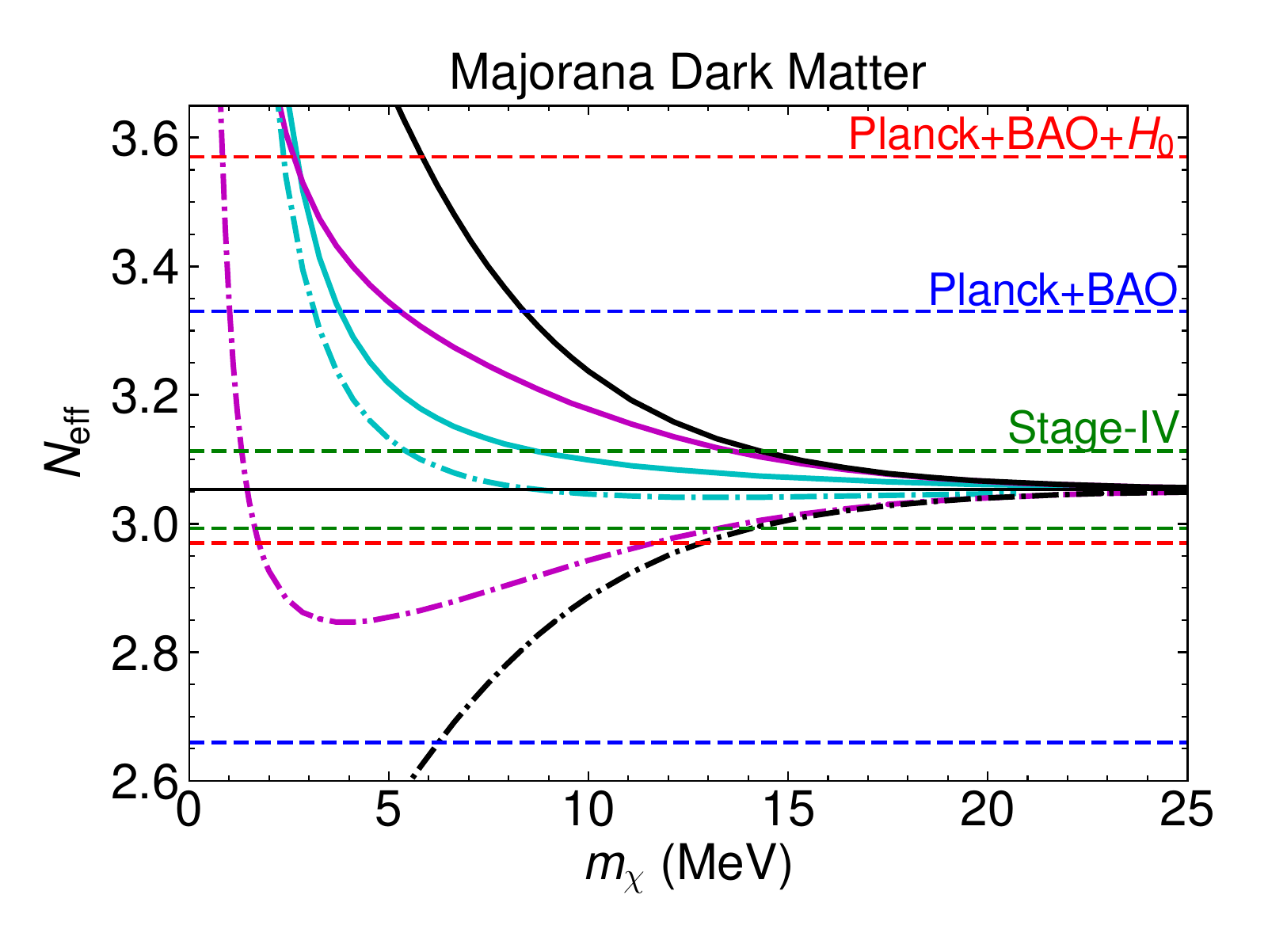} & \hspace{-0.7cm} \includegraphics[width=0.53\textwidth]{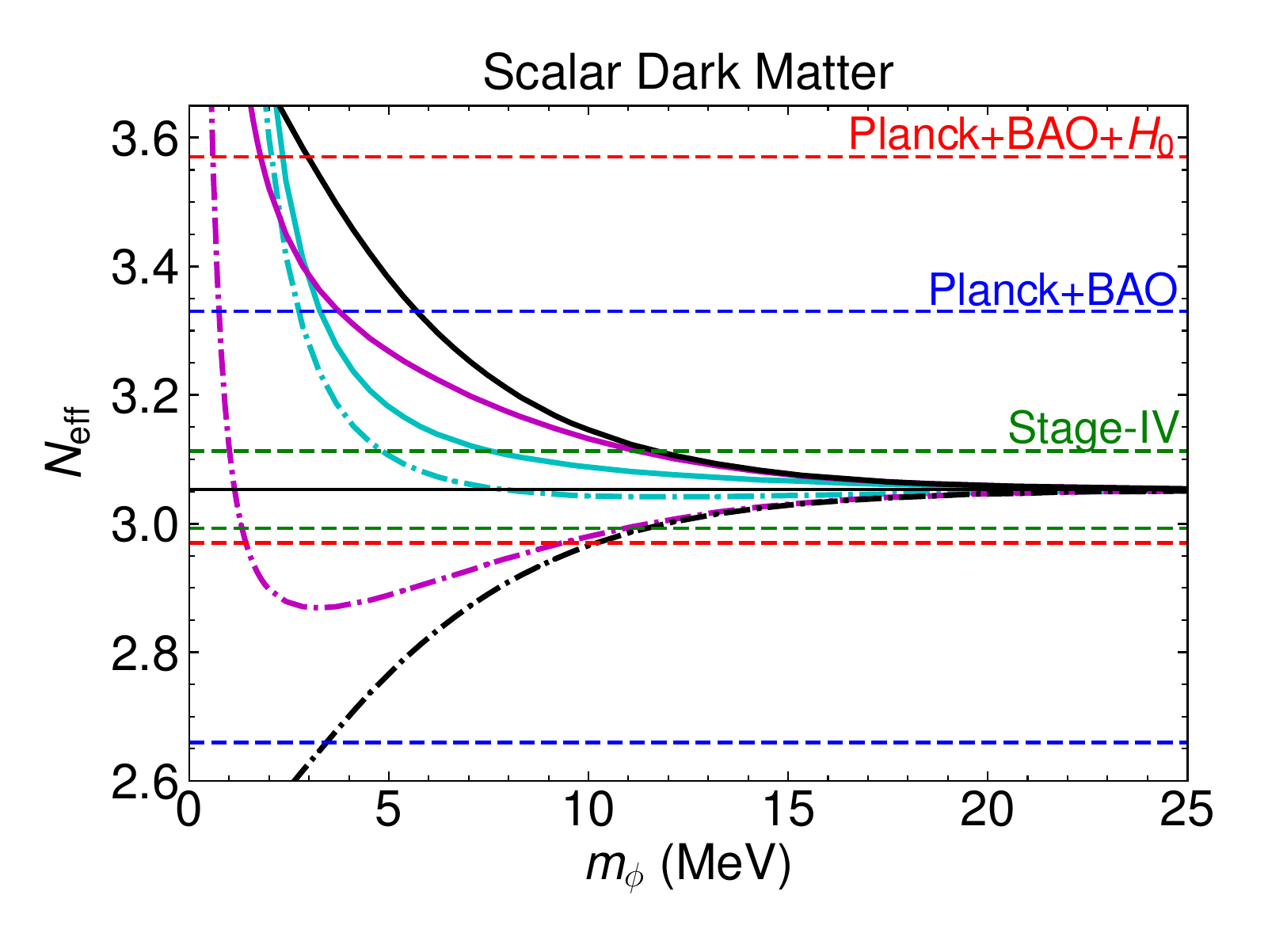}
\end{tabular}\vspace{-0.3cm}
\caption{\textit{Left panel:} $N_\text{eff}$ as relevant for CMB observations as a function of the mass of a Majorana thermal WIMP that annihilates to both electrons and neutrinos. \textit{Right panel:} same as the left panel but for a neutral scalar WIMP. The different lines correspond to different ratios of electrons and neutrinos in the annihilation final state. The solid lines correspond to a mostly electrophilic WIMP while the dashed-dotted correspond to mostly neutrinophilic one. Note that the calculated $N_{\rm eff}$ applies to both $s$-wave and $p$-wave annihilating relics.}\label{fig:DM_nu_and_e}
\end{figure}

\begin{table}[t]
\begin{center}
{\def\arraystretch{1.2}
\begin{tabular}{c|cc|cc|cc}
\hline\hline
   	 \multirow{2}{*}{Annihilation Ratio}    	 &   \multicolumn{6}{c}{$N_{\rm eff}$ Constraints } \\ \cline{2-7} 

   	 	 &   \multicolumn{2}{c|}{Planck}  &   \multicolumn{2}{c|}{Planck+BAO+$H_0$}  &   \multicolumn{2}{c}{Stage-IV Sensitivity}      \\ \hline
  	$e:\nu$			   &	$> 2.66$   &	$ < 3.33$	   	&  $>2.97$	   & $ < 3.57$   &  $> 2.99$ & $ <3.11$ \\ 
  \hline \hline
  $1:10^6$ &  -       &  8.4       &  -       &  5.8        &  -       &  14.3            \\ \hline
$1:10^5$ &  -       &  7.9       &  -       &  4.8        &  -       &  14.2           \\ \hline
$1:10^4$ &  -       &  5.3       &  -       &  2.6        &  -       &  13.6         \\ \hline
$1:10^3$ &  -       &  3.7        &  -       &  2.4        &  -       &  11.1         \\ \hline
$1:10^2$ &  -       &  3.8        &  -       &  2.7        &  -       &  8.7        \\ \hline
$1:9$ &  -       &  4.3        &  -       &  3.1        &  -       &  8.7        \\ \hline
$9:1$ &  -  &  4.0        &  -  &  3.0        & -  &  7.4          \\ \hline
$10^2:1$ &  -  &  3.1        &  -  &  2.4        &  -  &  5.4          \\ \hline
$10^3:1$ &  -    &  2.1       & -   &  1.6       &  -   &  3.2    \\ \hline
$10^4:1$ &  -    &  1.0        &  1.7-11.6   &  0.8        & 1.6-13.3 &  1.3       \\ \hline
$10^5:1$ &  0.6-5.0    &  -       &  0.3-12.7  &  -       &  0.3-14.0  &  0.2        \\ \hline
$10^6:1$ & 6.2    &  -       &  12.8  &  -       &  14.1  &  -         \\ \hline
 \hline
\end{tabular}
}
\end{center}\vspace{-0.3cm}
\caption{Lower bound on the mass of a generic Majorana WIMP (in MeV) at 95\% CL as a function of the annihilation final state ratio to electrons and neutrinos. The upper/lower rows correspond to electrophilic/neutrinophilic WIMPs. The only four slots with a range correspond to an exclusion range. Given the current disagreement between local and CMB measurements of the Hubble constant, a conservative bound will correspond to $N_{\rm eff}> 2.66$ and $N_{\rm eff}< 3.57$. Note that the bounds apply to both $s$-wave and $p$-wave annihilating relics. Constraints for other branching fractions can be inferred by interpolation or by use of the~\href{https://github.com/MiguelEA/nudec_BSM}{online code}.}\label{tab:SUMMARY_enu}
\end{table}

In Figure~\ref{fig:DM_nu_and_e} we show the resulting $N_{\rm eff}$, as relevant for CMB observations, for a Majorana and a scalar thermal dark matter particle for various annihilation final state ratios to electrons and neutrinos. In Table~\ref{tab:SUMMARY_enu} we provide a summary of the constraints -- that can be inferred from CMB observations -- for a generic Majorana dark matter particle that annihilates differently to electrons and neutrinos (which serves as a representative example of any dark matter particle). The reader is referred to Appendix~\ref{app:Tables} for tables with the constraints for any thermal dark matter particle, and figures of $N_{\rm eff}$ as a function of the WIMP mass.

We notice that unless the WIMP annihilates to electrons or neutrinos less than once every $10^{5}$ times, the resulting $N_{\rm eff}$ is quite different to that of the case in which the WIMP exclusively couples to either electrons or neutrinos, as can be appreciated from Figure~\ref{fig:DM_Neff}. In addition, the CMB bounds are milder for scenarios in which the WIMP couples non-negligibly to both electrons and neutrinos simultaneously. This is because, in such scenarios, neutrino decoupling is delayed as compared with the SM by the dark matter interactions. Which in turn means that the entropy released by the WIMP annihilation is shared between the neutrino and electromagnetic sectors, thus resulting in $N_{\rm eff} \sim N_{\rm eff}^{\rm SM}$. For the case of a Majorana dark matter particle, CMB observations rule out $m_\chi<0.6\,\text{MeV}$ at 95\% CL regardless of its annihilation final state ratio to electrons or neutrinos. While for a neutral scalar, CMB observations cannot probe $m_\phi > m_e$ independently of the annihilation final state. Although we expect that Stage-IV CMB experiments will generically improve the bound on the dark matter mass for any generic WIMP, we have found that there is a window of rather light WIMPs $m \in m_e-1.1 \,\text{MeV}$ with final state ratios to electrons and neutrinos of $\sim 10^4:1$ which will remain elusive even to high-sensitive CMB experiments. 

The bounds derived for WIMPs that annihilate to both electrons and neutrinos are relevant for various physically well motivated scenarios. For instance, a dark matter particle charged under a gauged $U(1)_{B-L}$ symmetry will annihilate to electrons and neutrinos at a similar rate~\cite{Okada:2010wd,Escudero:2018fwn}. Also, some neutrinophilic dark matter candidates do have small ($10^{-5}-10^{-4}$) but non-negligible annihilations to electrons, as in the case of dark matter within a $U(1)_{\mu -\tau}$ gauge symmetry~\cite{Arcadi:2018tly} or in scenarios with mixing between active and sterile neutrino states~\cite{Gonzalez-Macias:2016vxy}. Finally, on more general grounds, our bounds will be applicable to dark matter particles that annihilate through a $Z'$ with axial couplings to electrons or neutrinos~\cite{Langacker:2008yv}. An axial coupling to electrons implies the existence of an axial coupling to neutrinos, and the other way around, as granted by the SM particle content and $SU(2)_L$ gauge invariance.

\begin{table}[t]
\begin{center}
{\def\arraystretch{1.20}
\begin{tabular}{c|ccccc}
\hline\hline
   	 Annihilation Ratio   	 &   \multicolumn{5}{c}{$Y_p = 0.245 \pm 0.006$ (95\% CL) Constraints } \\ \hline
	$e:\nu$	 &  $\text{B},\, g=1$        &  $\text{B},\, g=2$         &  $\text{F},\, g=2$         &  $\text{B},\, g=3$        &  $\text{F},\, g=4$         \\
  \hline \hline
$1:10^6$ &  1.1        &  2.4        &  2.3        &  3.4        &  4.0        \\ \hline
$1:10^5$ &  1.1        &  2.0        &  2.0        &  2.4        &  2.7        \\ \hline
$1:10^4$ &  1.2        &  1.8        &  1.8        &  2.1        &  2.3        \\ \hline
$1:10^3$ &  1.4        &  1.8        &  1.8        &  2.0        &  2.2        \\ \hline
$1:10^2$ &  1.6        &  2.0        &  2.0        &  2.2        &  2.3        \\ \hline
$1:9$ &  1.9        &  2.2        &  2.2        &  2.4        &  2.5        \\ \hline
$9:1$ &  1.8        &  2.2        &  2.2        &  2.3        &  2.4        \\ \hline
$10^2:1$ &  1.5        &  1.9        &  1.9        &  2.1        &  2.2        \\ \hline
$10^3:1$ &  1.2        &  1.5        &  1.5        &  1.8        &  1.9        \\ \hline
$10^4:1$ &  0.7        &  1.1        &  1.1        &  1.4        &  1.6        \\ \hline
$10^5:1$ &  0.4        &  0.7        &  0.7        &  1.0        &  1.2        \\ \hline
$10^6:1$ &  0.4        &  0.5        &  0.5        &  0.7        &  0.9        \\ \hline
 \hline
\end{tabular}
}
\end{center}\vspace{-0.3cm}
\caption{Lower bound on the mass of a generic WIMP (in MeV) from the measurements of the primordial Helium abundance $Y_p$. The columns correspond to a neutral scalar, charged scalar, Majorana, neutral vector boson, and Dirac thermal dark matter particles. The rows correspond to different annihilation final states. }\label{tab:SUMMARY_enu_Yp}
\end{table}

\section{BBN constraints on thermal Dark Matter}\label{sec:BBN}
Light particles in thermal equilibrium with the SM plasma can impact the synthesis of the primordial elements during BBN~\cite{Kolb:1986nf,Serpico:2004nm,Boehm:2012gr,Boehm:2013jpa,Nollett:2014lwa}. The presence of such particles will affect BBN mainly because~\cite{Kolb:1986nf,Pospelov:2010hj,Iocco:2008va,Sarkar:1995dd} \textit{i)} the proton-to-neutron conversion rates are modified by a change in the neutrino-photon temperature evolution, and \textit{ii)} the early Universe expansion is modified, in particular the time at which Deuterium starts to form will be modified (while the temperature is fixed to be $T_\gamma \simeq 0.85\times10^{9}\,\text{K}$). 

In order to draw precise statements about the effect of MeV scale thermal dark matter particles on BBN, one should implement the temperature evolution obtained in Section~\ref{sec:BSM_neutrinodec} in a modern BBN code like \texttt{PArthENoPE}~\cite{Pisanti:2007hk,Consiglio:2017pot}, \texttt{AlterBBN}~\cite{Arbey:2011nf,Arbey:2018zfh} or \texttt{PRIMAT}~\cite{Pitrou:2018cgg}. However, that is beyond the scope of this work and will be left for future studies. Nonetheless, since the Helium abundance ($Y_p$) depends mainly upon the proton-to-neutron conversion rates and the time at which Deuterium forms~\cite{Pospelov:2010hj,Iocco:2008va,Sarkar:1995dd}, we can actually provide somewhat reliable estimates for $Y_p$ when generic WIMPs are present in the early Universe (see Appendix~\ref{app:Heliumabudances} for details about the calculation). 

\begin{figure}[t]
\centering
\hspace{1.4cm} \includegraphics[width=0.7\textwidth]{figures/legend}
\begin{tabular}{cc}
 \hspace{-0.8cm} \includegraphics[width=0.53\textwidth]{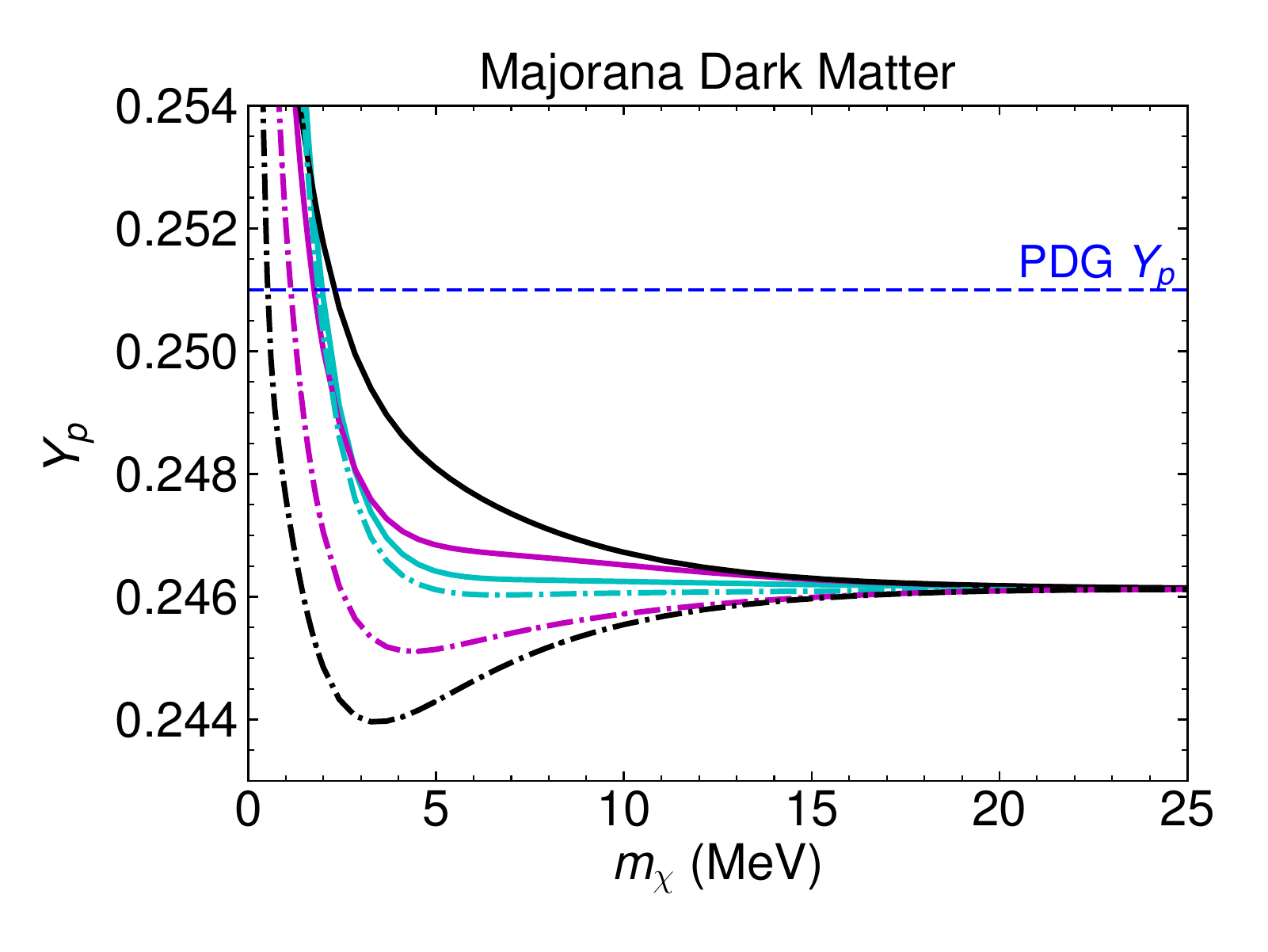}  & \hspace{-0.7cm} \includegraphics[width=0.53\textwidth]{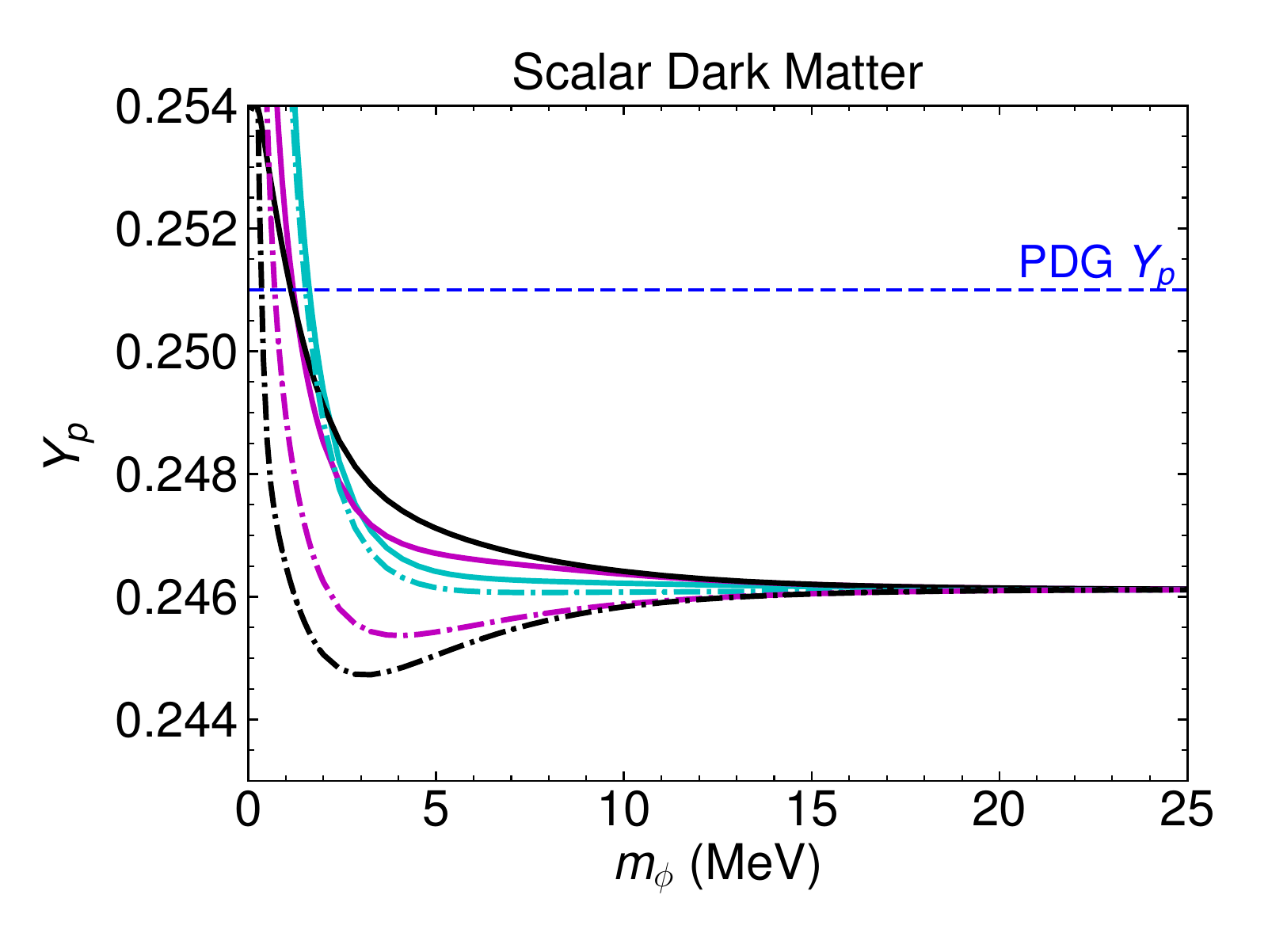} 
\end{tabular}\vspace{-0.3cm}
\caption{\textit{Left panel:} primordial Helium abundance as a function of the mass of a Majorana WIMP that annihilates to both electrons and neutrinos. \textit{Right panel:} same as the left panel but for a neutral scalar WIMP. The different lines correspond to different ratios of electrons and neutrinos in the annihilation final state. The solid lines correspond to a mostly electrophilic WIMP while the dashed-dotted correspond to a mostly neutrinophilic one. The horizontal dashed line corresponds to the 95\% CL observed upper limit on $Y_p$ as recommended from the PDG~\cite{pdg}. Note that the calculated values of $Y_p$ apply to both $s$-wave and $p$-wave annihilating relics.}\label{fig:DM_nu_and_e_Yp}
\end{figure}

In Figure~\ref{fig:DM_nu_and_e_Yp} we show the resulting Helium abundance as a function of the mass of a Majorana and a scalar WIMP depending on the annihilation final state ratio to electrons and neutrinos. In Table~\ref{tab:SUMMARY_enu_Yp} we show the resulting bounds on various dark matter particles by comparing the obtained Helium abundance to the measured value of $Y_p  = 0.245 \pm 0.06$ at 95\%CL as recommended by the PDG~\cite{pdg}. We notice that current measurements of the Helium abundance can be sensitive to WIMPs with masses $m < 0.4-4\,\text{MeV}$. Although we find that Helium BBN bounds are generically less constraining than those obtained from current CMB observations of $N_{\rm eff}$, the Helium BBN bounds can be slightly more constraining for WIMPs that annihilate to $e:\nu$ in ratios $\sim10^{4}:1$, thus providing complementary bounds to those inferred from CMB observations. 

\section{Summary, conclusions and outlook}\label{sec:summary}

In this study we have developed a simplified treatment of neutrino decoupling in the early Universe. Thanks to several well justified approximations we have been able to model the neutrino decoupling process in terms of two simple coupled differential equations for $T_\gamma$~\eqref{eq:T_gamma_QED} and $T_\nu$~\eqref{eq:dTnudt}. The virtues of our modeling are that: it incorporates the SM energy exchange rates between the electromagnetic and neutrino sectors of the plasma, it allows for a fast and precise evaluation of $N_{\rm eff}$, and it makes straightforward the inclusion of BSM species. This differs from previous studies in the literature that either assume that neutrinos decouple instantaneously, or that include all the relevant physics (exact collision terms, QED corrections, and neutrino oscillations), but which are computationally very expensive due to the complexity of the resulting kinetic equations.

In this study, we have illustrated the usefulness of our approach to neutrino decoupling in the early Universe to make precise statements about the value of $N_{\rm eff}$ beyond the Standard Model. In particular, we have studied the effect of thermal dark matter (WIMPs) on the process of neutrino decoupling. We have revisited the case of electrophilic and neutrinophilic thermal relics, and for the first time explored scenarios in which the WIMP annihilates simultaneously to electrons and neutrinos. For the latter case, we have shown that WIMPs -- in addition to releasing entropy into the system -- act as to generically delay the time of neutrino decoupling. By confronting the predicted values of $N_{\rm eff}$ from our treatment, with the observed values as reported by the Planck Collaboration in their 2018 analysis, we have set lower bounds on the masses of various types of WIMPs. Our analysis shows that Planck rules out both $s$-wave and $p$-wave electrophilic and neutrinophilic thermal dark matter candidates with $m < 3.0\,\text{MeV}$ at 95\% CL. We have pointed out that WIMPs with non-negligible interactions to both electrons and neutrinos are considerably more elusive to CMB observations than purely electrophilic and neutrinophilic dark matter particles. In addition, assisted by the precision of our approach to neutrino decoupling, we have shown that upcoming CMB observations will be sensitive to WIMPs of $m\lesssim 15\,\text{MeV}$. 

Although we have calculated the temperature evolution of the neutrino and electromagnetic sectors of the plasma for $T<20\,\text{MeV}$, we have not addressed in detail the impact of BSM sectors in the synthesis of the primordial element abundances during BBN. We have estimated the impact of light thermal dark matter on $Y_p$, and found that that particles with $m < 0.4-4\,\text{MeV}$ are expected to be constrained by current Helium abundance measurements. However, the actual implementation of the temperature evolution in a modern BBN code is beyond the scope of this work, and is left for future studies.  

Our approach to neutrino decoupling relies on very useful and well justified simplifying assumptions, which make the description of the neutrino decoupling process straightforward. Some improvements that would however lead to a more involved system of equations or non-analytical collision terms are: the inclusion of neutrino oscillations (which could be done with the methods developed in~\cite{Hannestad:2001iy}), including non-negligible chemical potentials for the neutrinos (which could be relevant for certain BSM scenarios), and to include spin statistics in the neutrino collision terms (which would require 2-d integrations at each time step). 

Finally, we note that the possibilities of our simplified approach to neutrino decoupling are countless given the variety of extensions of the Standard Model with massless or light particles (for instance, see~\cite{Escudero:2019gzq}). 

\section*{Acknowledgments}

The author thanks Dan Hooper, Gordan Krnjaic and Mathias Pierre for collaboration on related topics. The author is grateful to Sam Witte for a careful read of the draft, and also thanks Chris McCabe, Nuria Rius, Sergio Pastor and Sam Witte for their very helpful comments on earlier versions of the manuscript. The author is supported by the European Research Council under the European Union's Horizon 2020 program (ERC Grant Agreement No 648680 DARKHORIZONS).\\

This paper is dedicated to my mother (in memoriam), Antonia Abenza Ortiz, and to my partner, Andrea Gonz\'{a}lez Montoro, for their endless love, support and encouragement.

\appendix   
\section{Appendices} 
\subsection{Neutrino oscillations in the primordial Universe}\label{app:nuoscillations}
Here we provide an estimate for the temperature at which neutrinos begin to oscillate in the early Universe. The reader is referred to~\cite{Hannestad:2001iy,Dolgov:2002ab,Dolgov:2002wy} for detailed treatments. 

\begin{figure}[t]
\centering
\includegraphics[width=0.6\textwidth]{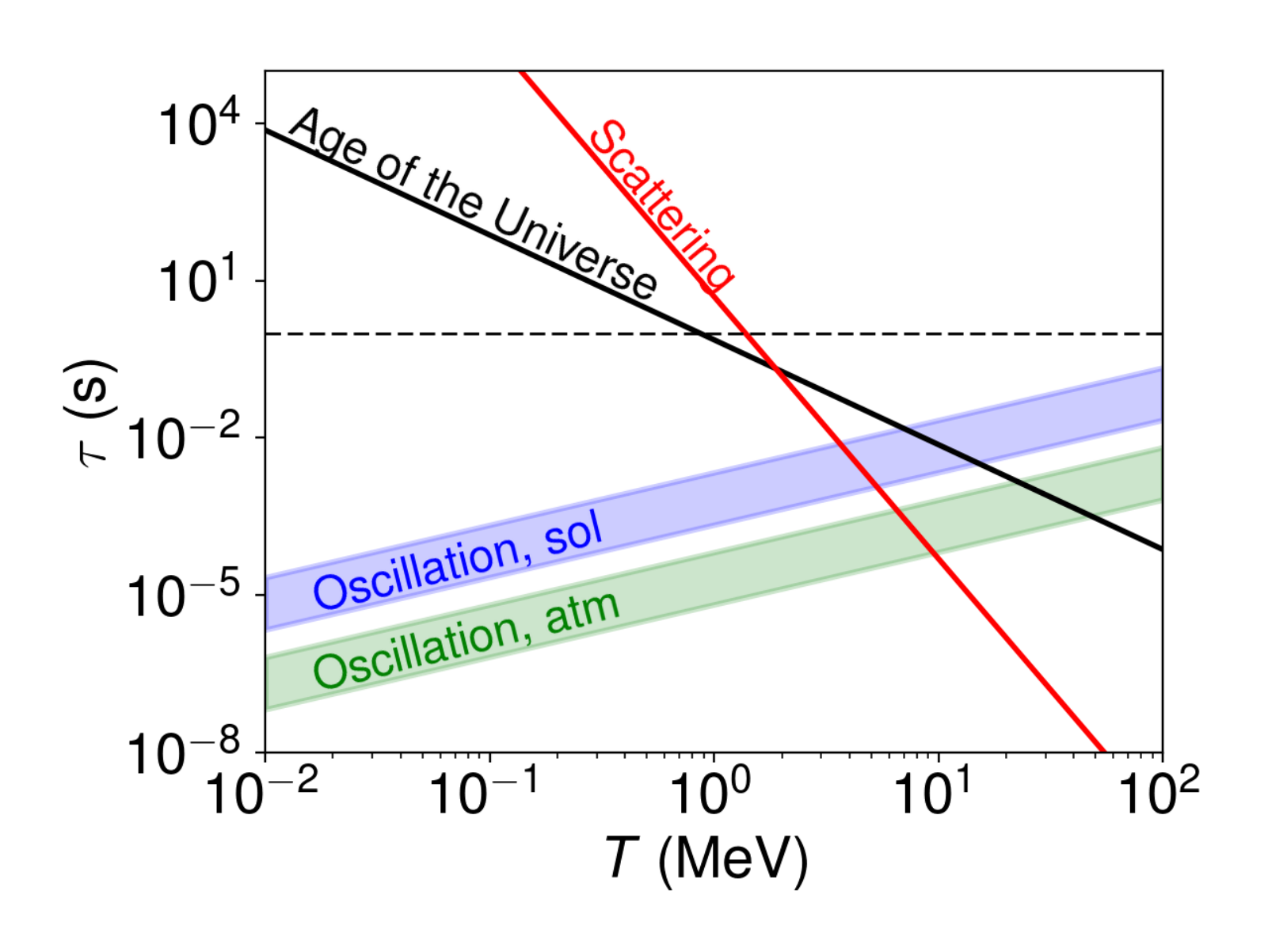}\vspace{-0.5cm} 
\caption{The timescales for the Universe's expansion, neutrino oscillation, and scattering interactions as relevant for neutrino decoupling. Since neutrino oscillations are active when $t_{\nu}^{\text{os}} < t_{\rm exp} < t_{\nu}^{\text{scat}}$, one can appreciate that this happens in the early Universe at $T\sim 3-5\,\text{MeV}$~\cite{Hannestad:2001iy,Dolgov:2002ab} given the observed pattern of masses and mixings~\cite{Esteban:2018azc,deSalas:2017kay,Capozzi:2016rtj}. }\label{fig:timescales}
\end{figure}

We can estimate the temperature at which neutrinos start to oscillate by comparing the characteristic time for oscillation, scattering, and expansion of the Universe. It is clear that neutrino oscillations will be active when $t_{\nu}^{\text{os}} < t_{\rm exp} < t_{\nu}^{\text{scat}}$, \textit{i.e.} when the oscillations take place within the lifetime of the Universe and are not decohered by scattering interactions. 
The oscillation timescale can easily be estimated from the 2 neutrino oscillation formula~\cite{pdg}:
\begin{align}
P(\nu_\alpha \to \nu_\beta, t) = \sin^2\theta_{\alpha\beta} \sin^2 \left[\frac{\Delta m^2}{4 E} \, c \, t \right] \, ,
\end{align}
where $E$ is the neutrino energy, $\Delta m^2$ is the mass squared difference, and $\theta_{\beta \alpha}$ corresponds to the mixing angle. Therefore, at the time at which the argument of the second sinus becomes $\mathcal{O}(1)$, the oscillations will be active unless the scattering interactions damp them. Taking $ E \to \left< E \right> = 3 \, T$ we can find all the relevant timescales:
 \begin{align}
t_{\nu}^{\text{os}} \sim \frac{12\,T}{\Delta m^2}\, , \qquad 
t_{\rm exp}  = \frac{1}{2H} \sim \frac{m_{Pl}}{3.44\sqrt{10.75} T^2} \, ,\qquad 
t_{\nu}^{\text{scat}} \sim \frac{1}{G_F^2 \,T^5} \, .
 \end{align}
Figure~\ref{fig:timescales} shows the evolution of the different timescales as relevant for neutrino decoupling. Hence, we can clearly appreciate that oscillations become active for $T\sim 3-5\,\text{MeV}$~\cite{Hannestad:2001iy,Dolgov:2002ab}. 

We also note that, even if BSM interactions are present, in order no to modify significantly $N_{\rm eff}$, the BSM-neutrino rates will become smaller than the Hubble rate at $T \sim\text{MeV}$ and therefore neutrino oscillations will be active regardless of possible BSM interactions given the strong energy dependence of the oscillation rate.

\subsection{Neutrino collision terms in the SM in the Maxwell-Boltzmann approximation}\label{app:collision_2body}
In this appendix we outline the results of the calculation of the collision terms for annihilations and scatterings assuming Maxwell-Boltzmann statistics for all the particles involved in the process, neglecting the statistical factors and assuming that all the species are massless. These results differ from those presented by the author in previous versions of the paper. The results outlined in previous versions of this work were based on the formulae presented in~\cite{Kawasaki:2000en} which we have found to have missing factors of 2 and 8 for the respective annihilation and scatterings interactions. Such errors have also been reported in footnote 4 of~\cite{Ichikawa:2005vw}.

We have been able to fully calculate the collision terms following the phase space reduction method developed in~\cite{Hannestad:1995rs}. We refer the reader to the appendices of~\cite{Fradette:2018hhl} and~\cite{Kreisch:2019yzn} which very recently appeared in the literature and which explain in more detail and with explicit formulae how to perform the different integration steps. 

By assuming Maxwell-Boltzmann statistics for all the particles involved in the process, neglecting the electron mass, neglecting statistical factors, and by considering the matrix elements reported in Table 2 of~\cite{Dolgov:2002wy} we arrive to the following annihilation collision term:
\begin{align}
\mathcal{C}_{\nu \bar{\nu} \leftrightarrow e^+e^-} (p) &= \frac{16 G_F^2 \left(g_L^2+g_R^2\right)}{3 \pi^3 }p \left[  T_\gamma^4 \, e^{-\frac{p}{T_\gamma}} - T_\nu^4 \, e^{-\frac{p}{T_\nu}} \right] \, , \label{eq:nu_e_ann_col}
\end{align}
where $G_F$ is the Fermi constant, $g_L^{\nu_e} = \frac{1}{2} + s_W^2$, $g_R^{\nu_e} = s_W^2$, $g_L^{\nu_{\mu,\,\tau}} = -\frac{1}{2} + s_W^2$, $g_L^{\nu_{\mu,\,\tau}} = s_W^2$, and $s_W$ is the sinus of the Weinberg angle. Expression~\eqref{eq:nu_e_ann_col} precisely matches the one obtained from Equation B.2 in~\cite{Dolgov:1997mb}.
The scattering rate between neutrinos and electrons can similarly be obtained to give:
\begin{align}
\mathcal{C}_{\nu e^\pm \leftrightarrow \nu e^\pm} (p) &= \frac{16 G_F^2 T_\gamma^2 \left(g_L^2+g_R^2\right)}{3 \pi ^3 p^2 (T_\nu-T_\gamma)^3} \left[-2 T_\gamma^2 e^{-\frac{p}{T_\nu}} \left(2 p^3 T_\gamma (T_\gamma-T_\nu)^3 +3 T_\nu^6 (T_\gamma+3 T_\nu)\right) \right. \\
&\left. +3 T_\nu^4 e^{-\frac{p}{T_\gamma}} (T_\gamma+3 T_\nu) \left(p^2 (T_\gamma-T_\nu)^2+2 p T_\gamma T_\nu (T_\nu-T_\gamma)+2 T_\gamma^2 T_\nu^2\right)  \right. \\
&\left. -3 (T_\gamma-T_\nu)^2 \left(T_\nu^4 \left(p^2+2 p T_\gamma+2 T_\gamma^2\right) e^{-p/T_\gamma}\right.\right. \\
&\left. \left. -T_\gamma^2 e^{-p/T_\nu} \left(p^3 (T_\gamma-T_\nu)+p^2 T_\nu^2+2 p T_\nu^3+2 T_\nu^4\right)\right)\right] \, .
\end{align}
If we were to consider only the forward part of the annihilation and scattering processes ($\bar{\nu} \nu \to {e^+ e^-},\,\nu e^\pm \to \nu e^\pm$) between electrons and neutrinos we will recover Equation 63 in~\cite{Dolgov:2002wy}. The momentum averaged rates are:
\begin{align}
\frac{1}{2\pi^2} \int p^3 \, \mathcal{C}_{\nu \bar{\nu} \leftrightarrow e^+e^-} \,  dp & = \frac{64 G_F^2 \left(g_L^2+g_R^2\right) }{\pi ^5} \, (T_\gamma^9-T_{\nu}^9)  \, , \label{eq:DeltaRho_ann}\\
\frac{1}{2\pi^2} \int p^2 \, \mathcal{C}_{\nu \bar{\nu} \leftrightarrow e^+e^-} \,  dp & =  \frac{16 G_F^2 \left(g_L^2+g_R^2\right) }{\pi ^5} \, (T_\gamma^8-T_{\nu}^8)\, , \label{eq:Deltan_ann}\\
\frac{1}{2\pi^2} \int p^3 \, \mathcal{C}_{\nu e^\pm \leftrightarrow \nu e^\pm} \,  dp & =  \frac{112 G_F^2 \left(g_L^2+g_R^2\right) }{\pi ^5}  \,  T_\gamma^4 \, T_\nu^4 \,(T_\gamma-T_\nu) \, .\label{eq:DeltaRho_scatt}
\end{align}
By following a similar procedure, we compute the neutrino-neutrino collision terms, and we perform the integration over momenta for all the relevant processes which yield:
\begin{align}
\frac{1}{2\pi^2} \int p^3 \, \mathcal{C}_{\nu_e \bar{\nu}_e \leftrightarrow \bar{\nu}\nu} \,  dp & =\frac{32 G_F^2 }{\pi ^5} \,(T_{\nu_\mu}^9-T_{\nu_e}^9) \, , \label{eq:DeltaRho_ann_nu}\\
\frac{1}{2\pi^2} \int p^2 \, \mathcal{C}_{\nu_e \bar{\nu}_e \leftrightarrow \bar{\nu}\nu} \,  dp & =  \frac{8 G_F^2 }{\pi ^5} \,(T_{\nu_\mu}^8-T_{\nu_e}^8) \, , \label{eq:DeltaRho_ann_nu}\\
\frac{1}{2\pi^2} \int p^3 \, \mathcal{C}_{\nu_e {\nu} \leftrightarrow {\nu_e} \nu} \,  dp &= \frac{48 G_F^2}{\pi ^5} \, T_{\nu_e}^4 \, T_{\nu_\mu}^4 \, (T_{\nu_\mu}-T_{\nu_e})   \, , \label{eq:Deltan_scatt_nu_1}\\
\frac{1}{2\pi^2} \int p^3 \, \mathcal{C}_{\nu_e \bar{\nu} \leftrightarrow \nu_e\bar{\nu}} \,  dp & = \frac{8 G_F^2 }{\pi ^5}\,  T_{\nu_e}^4 \, T_{\nu_\mu}^4 \, (T_{\nu_\mu}-T_{\nu_e}) \, . \label{eq:Deltan_scatt_nu_2}
\end{align}
With these results, we are able to write the total number density and energy density transfer rates in the SM within the Maxwell-Boltzmann approximation:
\begin{subequations}\label{eq:energyrates_nu_SM_MB}
\begin{align}
\left. \frac{\delta \rho_{\nu_e}}{\delta t} \right|_{\rm SM}^{\rm MB} &= \frac{G_F^2}{\pi^5}\left[\left( 1 + 4 s_W^2 + 8 s_W^4 \right) \, F_{\rm MB}(T_\gamma,T_{\nu_e}) + 2 \, F_{\rm MB}(T_{\nu_\mu},T_{\nu_e}) \right] \, ,\\
\left. \frac{\delta \rho_{\nu_\mu}}{\delta t} \right|_{\rm SM}^{\rm MB} &= \frac{G_F^2}{\pi^5}\left[\left( 1 - 4 s_W^2 + 8 s_W^4 \right) \, F_{\rm MB}(T_\gamma,T_{\nu_\mu}) -   F_{\rm MB}(T_{\nu_\mu},T_{\nu_e}) \right] \, ,
\end{align}
\end{subequations}
where we have defined:
\begin{align}
F_{\rm MB}(T_1,T_2) = 32 \, (T_1^9-T_2^9) + 56 \, T_1^4\,T_2^4 \, (T_1-T_2)\, .
\end{align}
Similarly, the number density transfer rates read:
\begin{subequations}
\begin{align}
\left. \frac{\delta n_{\nu_e}}{\delta t} \right|_{\rm SM}^{\rm MB} &= 8\frac{G_F^2}{\pi^5}\left[\left( 1 + 4 s_W^2 + 8 s_W^4 \right) \, (T_\gamma^8 - T_{\nu_e}^8) + 2 \, (T_{\nu_\mu}^8 - T_{\nu_e}^8) \right] \, ,\\
\left. \frac{\delta n_{\nu_\mu}}{\delta t} \right|_{\rm SM}^{\rm MB} &= 8\frac{G_F^2}{\pi^5}\left[\left( 1 - 4 s_W^2 + 8 s_W^4 \right) \, (T_\gamma^8 - T_{\nu_\mu}^8) -  (T_{\nu_\mu}^8 - T_{\nu_e}^8) \right] \, ,
\end{align}
\end{subequations}
where $\delta n_i/\delta t \equiv \sum_j \left<\sigma v\right>_{ij} (n_i^2-n_j^2)$.

\subsection{Thermodynamics formulae}
Here we list the formulae used to compute the energy density, pressure density, and the variation of the energy density with respect to the temperature for the relevant species:
\begin{align}
& \rho_{\nu} = 2\, \frac{7}{8} \frac{\pi^2}{30} T^4 \, , \qquad \partial \rho_\nu/\partial T = 4 \rho_\nu/T \, , \qquad  
\rho_{\gamma} =2 \,  \frac{\pi^2}{30} T^4 \, , \qquad \partial \rho_\gamma/\partial T = 4 \rho_\gamma/T \, ,  \\
&\rho_{e} = \frac{4}{2\pi^2} \int_{m_e}^{\infty}  dE \,  \frac{E^2 \sqrt{E^2-m_e^2}}{e^{E/T} +1 } \, ,\qquad 
p_{e} = \frac{4}{6\pi^2} \int_{m_e}^{\infty}  dE \,  \frac{(E^2-m_e^2)^{3/2}}{e^{E/T} +1 } \, ,\\
 &  \partial \rho_e/\partial T = \frac{4}{2\pi^2} \int_{m_e}^{\infty}  dE \,  \frac{E^3 \sqrt{E^2-m_e^2}}{4 T^2} \cosh^{-2}\left[\frac{E}{2T}\right] \,, \\
 &\rho_{\chi}^{\rm F} = \frac{g_\chi}{2\pi^2} \int_{m_\chi}^{\infty}  dE \,  \frac{E^2 \sqrt{E^2-m_\chi^2}}{e^{E/T} +1 } \, ,\qquad 
p_{\chi}^{\rm F} = \frac{g_\chi}{6\pi^2} \int_{m_\chi}^{\infty}  dE \,  \frac{(E^2-m_\chi^2)^{3/2}}{e^{E/T} +1 } \, ,\\
 &  \partial \rho_\chi^{\rm F}/\partial T = \frac{g_\chi}{2\pi^2} \int_{m_\chi}^{\infty}  dE \,  \frac{E^3 \sqrt{E^2-m_\chi^2}}{4 T^2} \cosh^{-2}\left[\frac{E}{2T}\right] \,, \\
  &\rho_{\chi}^{\rm B} = \frac{g_\chi}{2\pi^2} \int_{m_\chi}^{\infty}  dE \,  \frac{E^2 \sqrt{E^2-m_\chi^2}}{e^{E/T} -1 } \, ,\qquad 
p_{\chi}^{\rm B} = \frac{g_\chi}{6\pi^2} \int_{m_\chi}^{\infty}  dE \,  \frac{(E^2-m_\chi^2)^{3/2}}{e^{E/T} -1 } \, ,\\
 &  \partial \rho_\chi^{\rm B}/\partial T = \frac{g_\chi}{2\pi^2} \int_{m_\chi}^{\infty}  dE \,  \frac{E^3 \sqrt{E^2-m_\chi^2}}{4 T^2} \sinh^{-2}\left[\frac{E}{2T}\right] \,.
\end{align}
Note that in practice, we truncate the integral for massive particles at $E = 100\,T$, and set $\rho_\chi = p_\chi = \partial\rho_\chi/\partial T = 0$ for $T< m_\chi/30$. 
\subsection{Helium abundance}\label{app:Heliumabudances}
Here we outline the equations used for our calculation of the primordial Helium abundance. The Boltzmann equation that governs the number of neutrons in the early Universe is~\cite{Sarkar:1995dd}:
\begin{align}
\frac{dX_n}{dt} = \Gamma_{pn} (1-X_n) - \Gamma_{np} \, X_n \, ,
\end{align}
where $X_n \equiv \frac{n_n}{n_n+n_p}$. We can easily estimate the Helium abundance by noting that nearly all the present neutrons at the time at which Deuterium is no longer dissociated by photons ($T_D = 0.85 \times 10^{9}\,\text{K} = 0.073\,\text{MeV}$) will form Helium~\cite{Sarkar:1995dd}. This means that:
\begin{align}
Y_p  \simeq {2 X_n}|_{T=T_{D}} \, .
\end{align}
In the early Universe, there are three processes that can efficiently convert protons into neutrons and the other way around. They are $n+\nu_e \leftrightarrow p+e^-$, $n+e^+ \leftrightarrow p + \bar{\nu}_e$ and $n \leftrightarrow p + e^- + \bar{\nu}_e$. The rate for neutron to proton conversion reads~\cite{Dicus:1982bz}:
\begin{eqnarray}\nonumber
\Gamma_{np} = K\, \int_1^\infty  d\epsilon \frac{(\epsilon -q)^2 (\epsilon^2-1)^{1/2} \epsilon }{(1+e^{-\epsilon z_\gamma})(1+e^{(\epsilon-q)z_\nu})} +  K\, \int_1^\infty  d\epsilon \frac{(\epsilon +q)^2 (\epsilon^2-1)^{1/2} \epsilon }{(1+e^{\epsilon z_\gamma})(1+e^{-(\epsilon+q)z_\nu})} \, ,
\end{eqnarray}
where $z_\gamma = m_e/T_\gamma$, $z_\nu=m_e/T_\nu$, $q = (m_n - m_p)/m_e $, and $m_n - m_p = 1.2933\,\text{MeV}$~\cite{pdg} and $ K \simeq(1.636 \, \tau_n)^{-1}$ and we take $\tau_n =880.2\,\text{s}$~\cite{pdg}. The proton-to-neutron rates are obtained by $\Gamma_{pn} = \Gamma_{np}(-q)$.
\subsection{Supplemental material for generic thermal Dark Matter candidates}\label{app:Tables}
Here we show the corresponding $N_{\rm eff}$ as a function of the mass of a generic Majorana, Dirac, neutral scalar, charged scalar, and neutral vector boson dark matter particle. We also provide tables with the corresponding bounds given the observed values of $N_{\rm eff}$ from current CMB observations and the sensitivity from upcoming CMB experiments. Constraints for other branching fractions can be inferred by interpolation or by use of the~\href{https://github.com/MiguelEA/nudec_BSM}{online code}.

\begin{figure}[t]
\centering
 \includegraphics[width=0.97\textwidth]{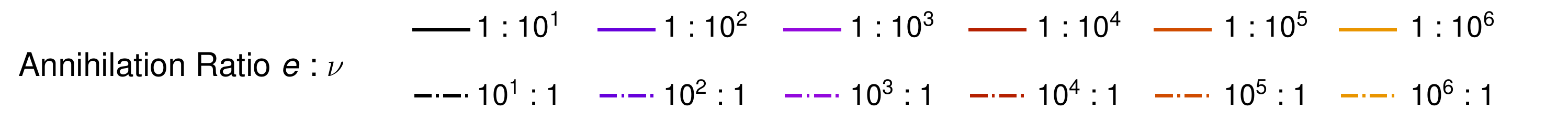}
\begin{tabular}{cc}
\hspace{-0.8cm} \includegraphics[width=0.53\textwidth]{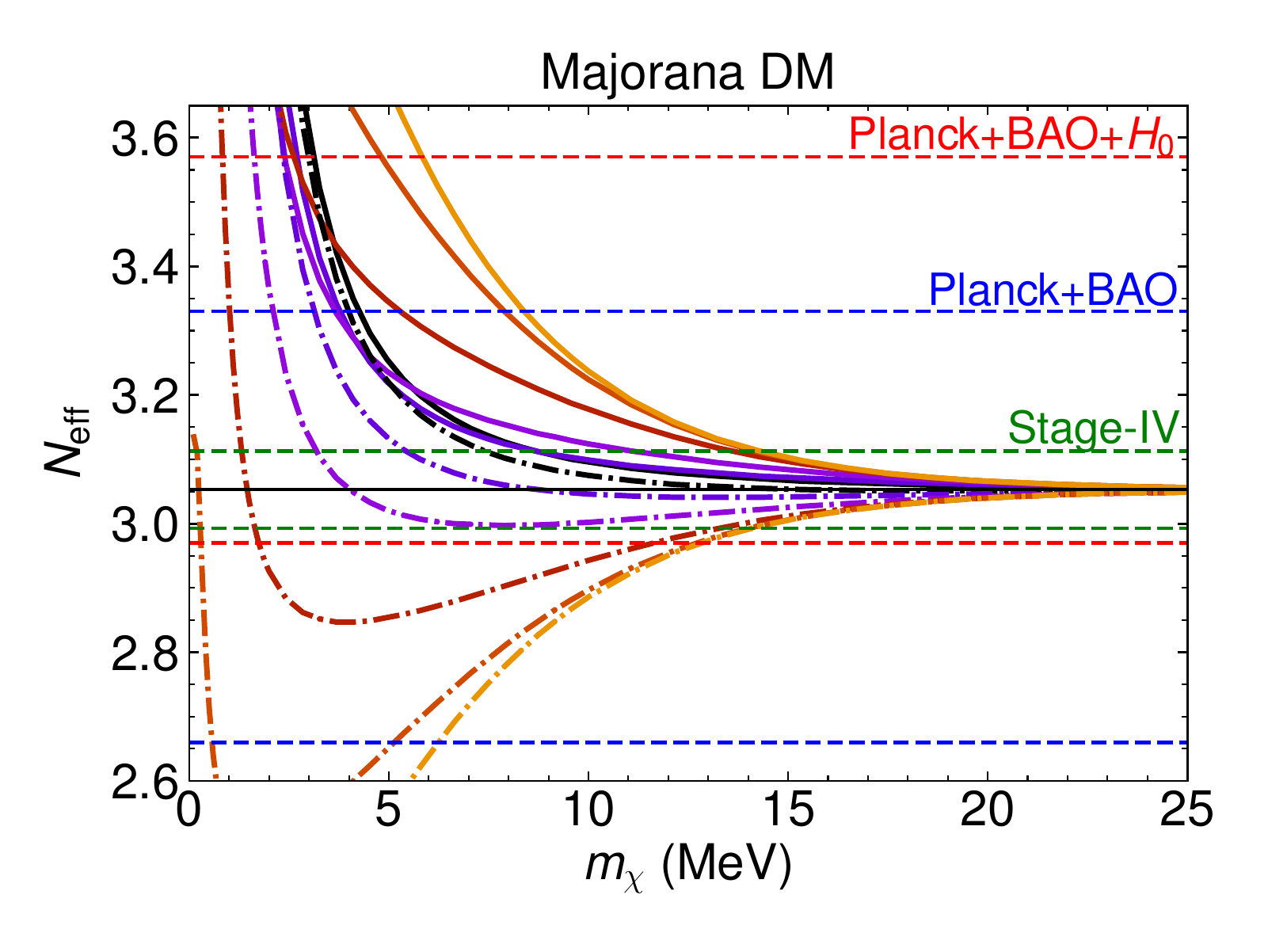} & \hspace{-0.7cm} \includegraphics[width=0.53\textwidth]{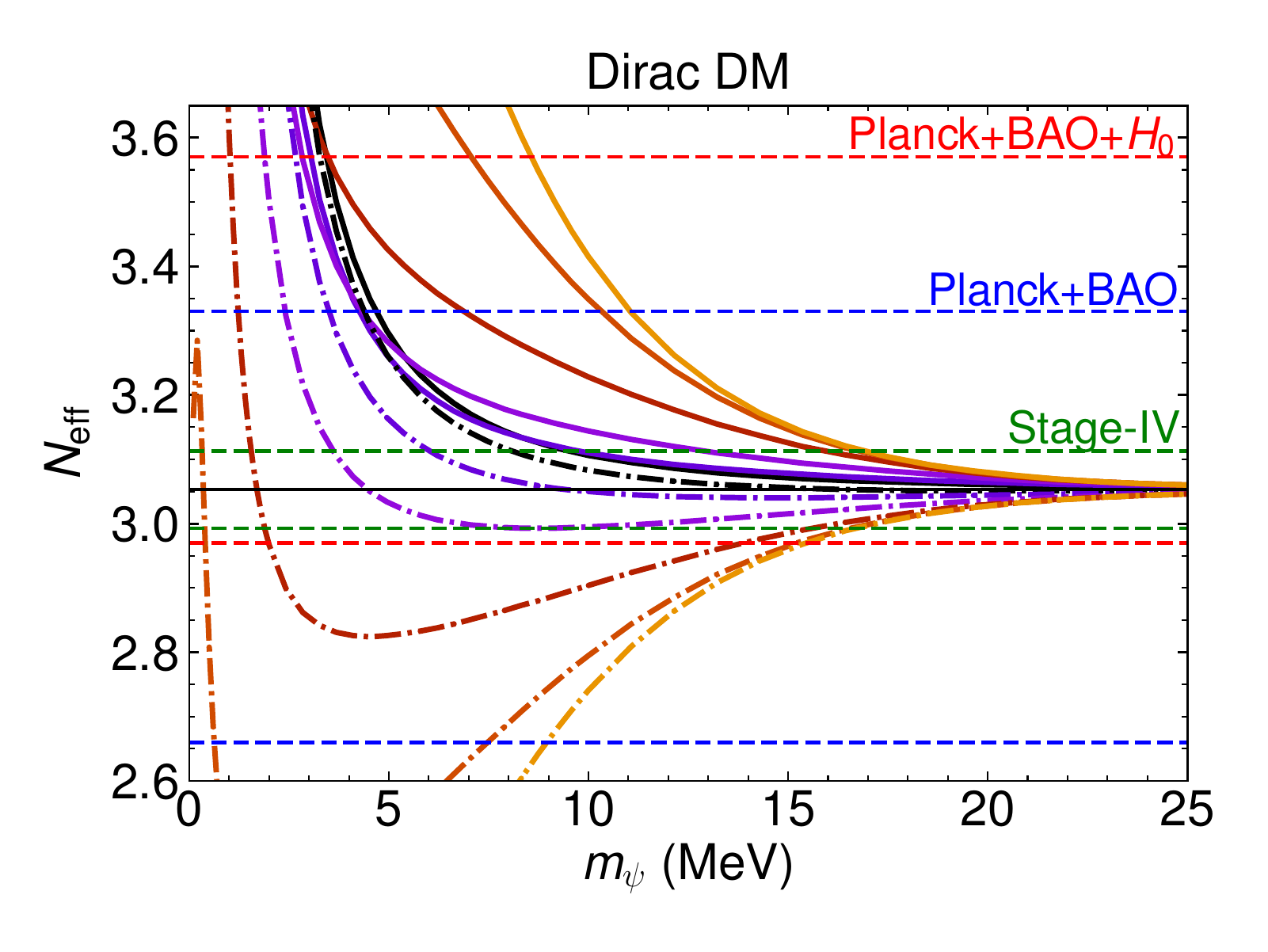} \\
\hspace{-0.8cm} \includegraphics[width=0.53\textwidth]{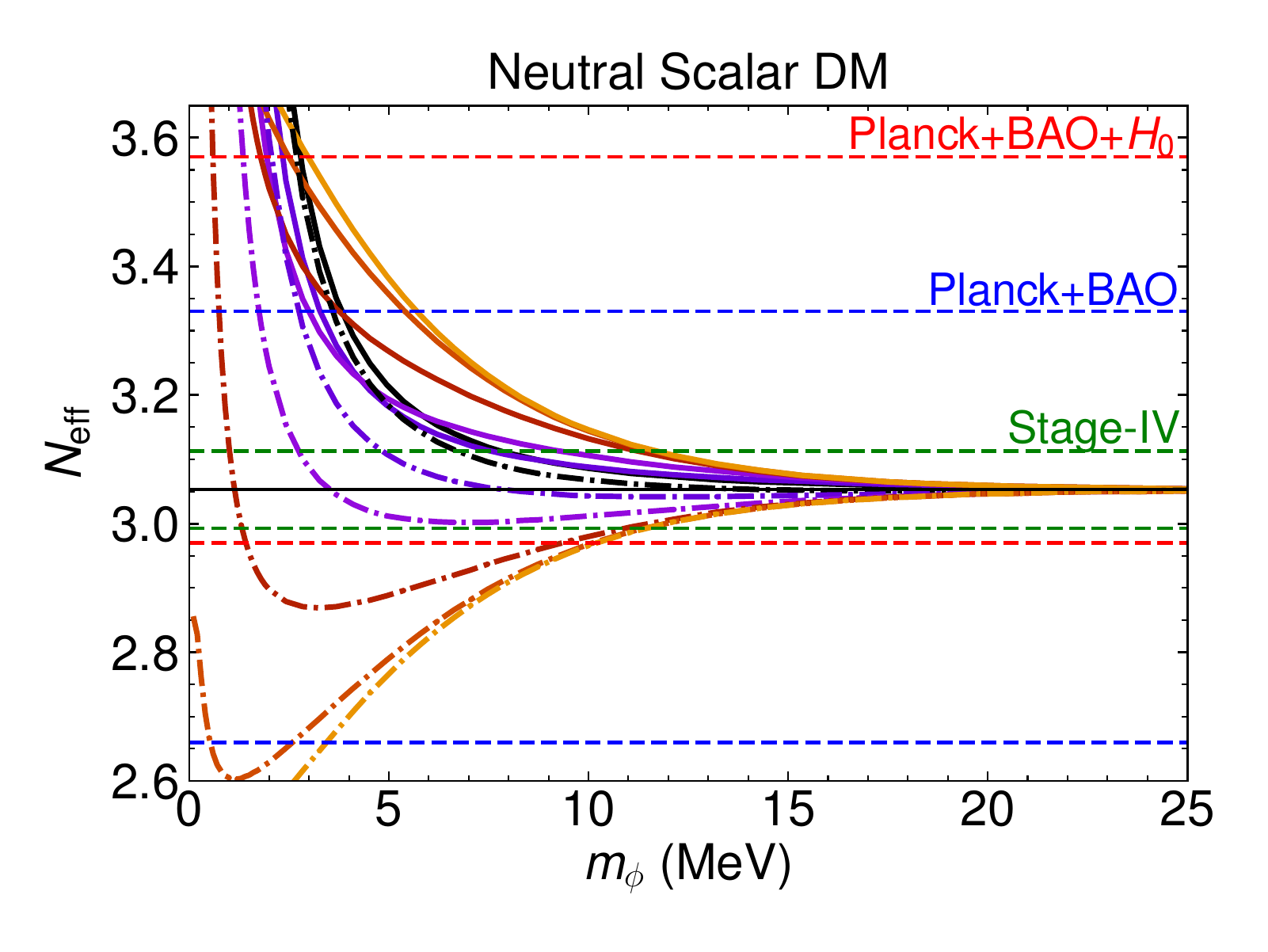} & \hspace{-0.7cm} \includegraphics[width=0.53\textwidth]{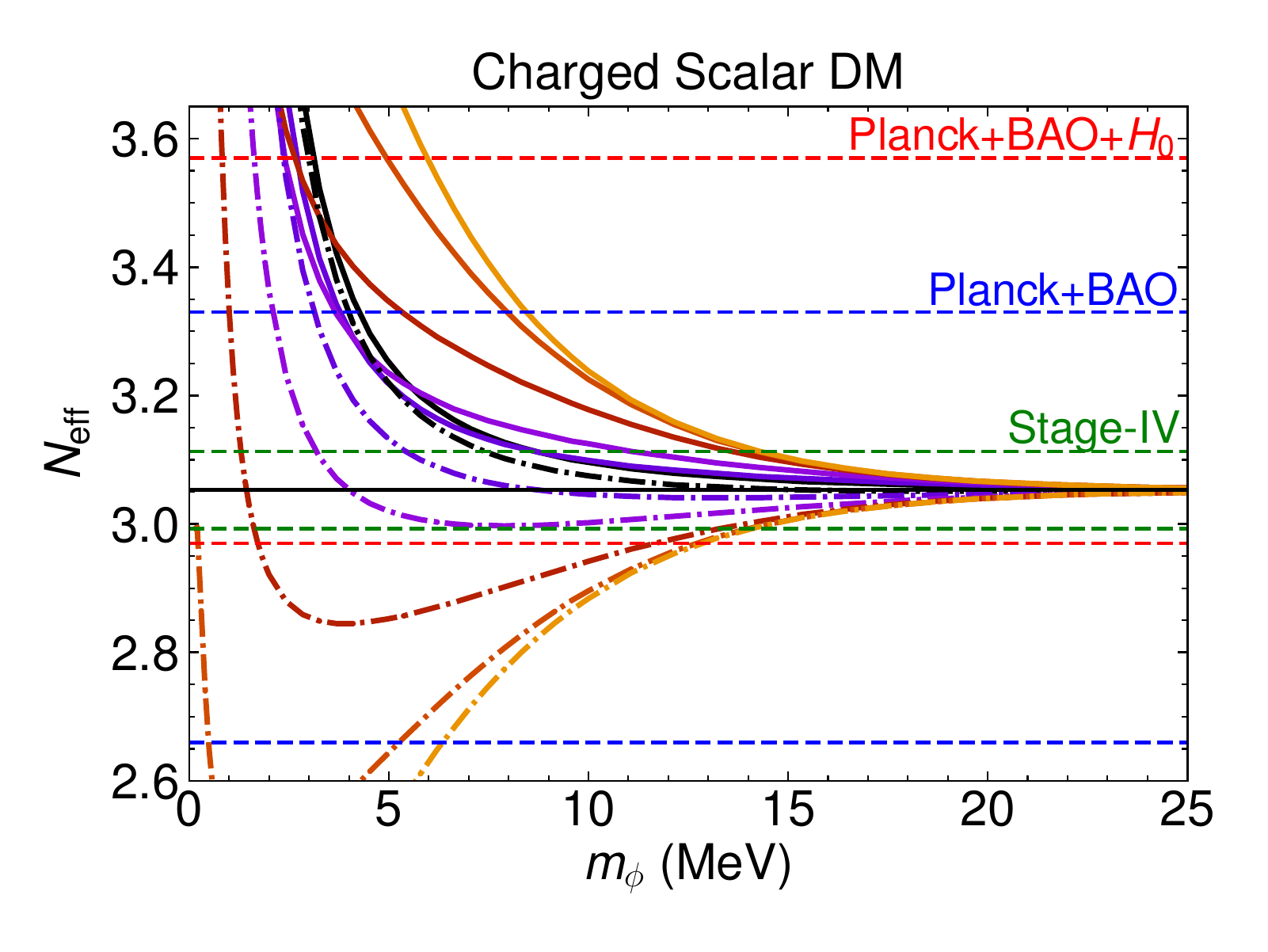}
\end{tabular}\vspace{-0.2cm}
\includegraphics[width=0.53\textwidth]{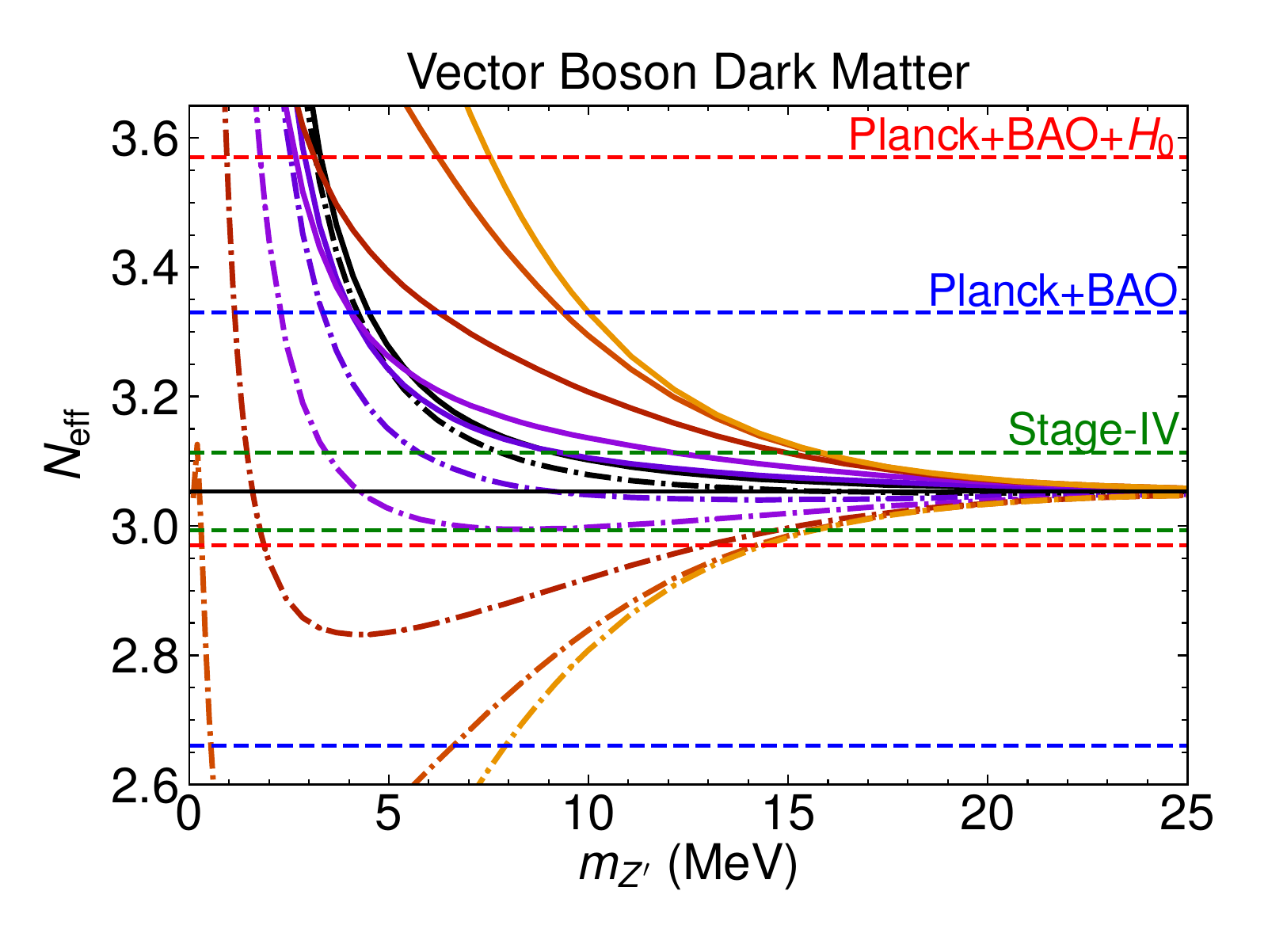} \vspace{-0.2cm}
\caption{$N_\text{eff}$ as relevant for CMB observations as a function of the mass of various thermal dark matter particles that annihilate to both electrons and neutrinos. The solid lines correspond to mostly electrophilic WIMPs while the dashed-dotted correspond to mostly neutrinophilic WIMPs. Note that the calculated $N_{\rm eff}$ applies to both $s$-wave and $p$-wave annihilating relics. }\label{fig:DM_nu_and_e_app}
\end{figure}

\begin{table}[h]
\begin{center}
{\def\arraystretch{1.0}
\begin{tabular}{c|cc|cc|cc}
\hline\hline
   	 \multirow{2}{*}{Annihilation Ratio}    	 &   \multicolumn{6}{c}{$N_{\rm eff}$ Constraints } \\ \cline{2-7} 

   	 	 &   \multicolumn{2}{c|}{Planck+BAO}  &   \multicolumn{2}{c|}{Planck+BAO+$H_0$}  &   \multicolumn{2}{c}{Stage-IV Sensitivity}      \\ \hline

  	$e:\nu$			   &	$> 2.66$   &	$ < 3.33$	   	&  $>2.97$	   & $ < 3.57$   &  $ >2.975$ & $ <3.095$ \\ 
  \hline \hline
  $1:10^6$ &  -       &  5.7        &  -          &  3.0        &  -          &  11.7       \\ \hline
  $1:10^5$ &  -       &  5.4        &  -          &  2.5        &  -          &  11.6       \\ \hline
$1:10^4$ &  -       &  3.8        &  -          &  1.8        &  -          &  11.2       \\ \hline
$1:10^3$ &  -       &  3.0        &  -          &  1.9        &  -          &  9.3	\\ \hline
$1:10^2$ &  -       &  3.3        &  -          &  2.3        &  -          &  7.5 	\\ \hline
$1:9$ &  -       &  3.8        &  -          &  2.8        &  -          &  7.8        \\ \hline
$9:1$ &  -   &  3.6        &  -          &  2.7        &  -          &  6.7        \\ \hline
$10^2:1$ &  -  &  2.8        &  -          &  2.1        &  -          &  4.8        \\ \hline
$10^3:1$ &  -    &  1.7        &  -          &  1.3      &   -          &  2.7        \\ \hline
$10^4:1$ &  -    &  0.7        &  1.4-9.3          &  0.6        &  1.3-10.9          &  1.0        \\ \hline
$10^5:1$ &  0.5-2.6          &  -          &  10.1        &  -          &  11.4       &  -          \\ \hline
$10^6:1$ & 3.4        &  -          &  10.2        &  -          &  11.5       &  -          \\ \hline
 \hline
\end{tabular}
}
\end{center}\vspace{-0.5cm}
\caption{Same as Table~\ref{tab:SUMMARY_enu} but for a neutral scalar boson dark matter particle.}\label{tab:SUMMARY_enu_SCALAR}
\end{table}
\vspace{-0.5cm}
\begin{table}[h]
\begin{center}
{\def\arraystretch{1.0}
\begin{tabular}{c|cc|cc|cc}
\hline\hline
   	 \multirow{2}{*}{Annihilation Ratio}    	 &   \multicolumn{6}{c}{$N_{\rm eff}$ Constraints } \\ \cline{2-7} 

   	 	 &   \multicolumn{2}{c|}{Planck+BAO}  &   \multicolumn{2}{c|}{Planck+BAO+$H_0$}  &   \multicolumn{2}{c}{Stage-IV Sensitivity}      \\ \hline

  	$e:\nu$			   &	$> 2.66$   &	$ < 3.33$	   	&  $>2.97$	   & $ < 3.57$   &  $ >2.975$ & $ <3.095$ \\ 
  \hline \hline
  $1:10^6$ &  -       &  8.4        &  -          &  6.0        &  -          &  14.3       \\ \hline
  $1:10^5$ &  -       &  8.0        &  -          &  4.9        &  -          &  14.2       \\ \hline
$1:10^4$ &  -       &  5.3        &  -          &  2.6        &  -          &  13.6       \\ \hline
$1:10^3$ &  -       &  3.7        &  -          &  2.4        &  -          &  11.1	\\ \hline
$1:10^2$ &  -       &  3.8        &  -          &  2.7        &  -          &  8.7 	\\ \hline
$1:9$ &  -       &  4.3        &  -          &  3.1       &  -          &  8.7        \\ \hline
$9:1$ &  -   &  4.0        &  -          &  3.0        &  -          &  7.4        \\ \hline
$10^2:1$ &  -  &  3.1        &  -          &  2.4        &  -          &  5.4        \\ \hline
$10^3:1$ &  -    &  2.1        &   -       &    1.4   &  -          &  3.2        \\ \hline
$10^4:1$ &  -    &  1.0        &  1.7-11.7          &  0.6        &  1.6-13.3          &  1.3        \\ \hline
$10^5:1$ &  0.5-5.3          &  -          &  18.0        &  -          &  14.0       &  -          \\ \hline
$10^6:1$ &  6.3        &  -          &  12.9        &  -          &  14.1       &  -          \\ \hline
 \hline
\end{tabular}
}
\end{center}\vspace{-0.5cm}
\caption{Same as Table~\ref{tab:SUMMARY_enu} but for a charged scalar boson dark matter particle.}\label{tab:SUMMARY_enu_SCALAR_CHARGED}
\end{table}

\begin{table}[h]
\begin{center}
{\def\arraystretch{1.23}
\begin{tabular}{c|cc|cc|cc}
\hline\hline
   	 \multirow{2}{*}{Annihilation Ratio}    	 &   \multicolumn{6}{c}{$N_{\rm eff}$ Constraints } \\ \cline{2-7} 

   	 	 &   \multicolumn{2}{c|}{Planck+BAO}  &   \multicolumn{2}{c|}{Planck+BAO+$H_0$}  &   \multicolumn{2}{c}{Stage-IV Sensitivity}      \\ \hline

  	$e:\nu$			   &	$> 2.66$   &	$ < 3.33$	   	&  $>2.97$	   & $ < 3.57$   &  $ >2.975$ & $ <3.095$ \\ 
  \hline \hline
  $1:10^6$ &  -       &  10.0        &  -          &  7.5        &  -          &  15.9       \\ \hline
  $1:10^5$ &  -       &  9.4        &  -          &  6.2        &  -          &  15.8       \\ \hline
$1:10^4$ &  -       &  6.2        &  -          &  3.1        &  -          &  15.0       \\ \hline
$1:10^3$ &  -       &  4.1        &  -          &  2.7        &  -          &  12.1	\\ \hline
$1:10^2$ &  -       &  4.1        &  -          &  2.8        &  -          &  9.3 	\\ \hline
$1:9$ &  -       &  4.5        &  -          &  3.3        &  -          &  9.2        \\ \hline
$9:1$ &  -   &  4.2        &  -          &  3.2        &  -          &  7.8        \\ \hline
$10^2:1$ &  -  &  3.3        &  -          &  2.5        &  -          &  5.8        \\ \hline
$10^3:1$ &  -    &  2.3        &   -        &    1.8   &   3.7-16.6         &  3.4        \\ \hline
$10^4:1$ &  -    &  1.1        &  1.9-13.0           &  0.9        & 1.8-14.7          &  1.4        \\ \hline
$10^5:1$ &  0.5-6.6          &  -          &  0.3-14.2        &  -          &  0.3-15.5       &  -          \\ \hline
$10^6:1$ &  7.9        &  -          &  14.4        &  -          &  15.6       &  -          \\ \hline
 \hline
\end{tabular}
}
\end{center}
\caption{Same as Table~\ref{tab:SUMMARY_enu} but for a neutral vector boson dark matter particle.}\label{tab:SUMMARY_enu_Zp}
\end{table}

\begin{table}[h]
\begin{center}
{\def\arraystretch{1.23}
\begin{tabular}{c|cc|cc|cc}
\hline\hline
   	 \multirow{2}{*}{Annihilation Ratio}    	 &   \multicolumn{6}{c}{$N_{\rm eff}$ Constraints } \\ \cline{2-7} 

   	 	 &   \multicolumn{2}{c|}{Planck+BAO}  &   \multicolumn{2}{c|}{Planck+BAO+$H_0$}  &   \multicolumn{2}{c}{Stage-IV Sensitivity}      \\ \hline

  	$e:\nu$			   &	$> 2.66$   &	$ < 3.33$	   	&  $>2.97$	   & $ < 3.57$   &  $ >2.975$ & $ <3.095$ \\ 
  \hline \hline
  $1:10^6$ &  -       &  11.0        &  -          &  8.6        &  -          &  16.9       \\ \hline
  $1:10^5$ &  -       &  10.3        &  -          &  7.1        &  -          &  16.8       \\ \hline
$1:10^4$ &  -       &  6.9        &  -          &  3.5        &  -          &  15.9       \\ \hline
$1:10^3$ &  -       &  4.3        &  -          &  2.8        &  -          &  12.9	\\ \hline
$1:10^2$ &  -       &  4.3        &  -          &  3.1        &  -          &  9.7 	\\ \hline
$1:9$ &  -       &  4.7        &  -          &  3.4        &  -          &  9.5        \\ \hline
$9:1$ &  -   &  4.4        &  -          &  3.3        &  -          &  8.1        \\ \hline
$10^2:1$ &  -  &  3.5        &  -          &  2.7        &  -          &  6.0        \\ \hline
$10^3:1$ &  -    &  2.4        &  -          &  1.9      &  -        &  3.6        \\ \hline
$10^4:1$ & -    &  0.6        &  2.0-13.9          &  1.0        &  1.9-15.6          &  1.6        \\ \hline
$10^5:1$ &  0.6-7.5          &  -          &  0.4-15.2        &  -          &  0.4-16.6       &  0.3          \\ \hline
$10^6:1$ &  9.0        &  -          &  15.5        &  -          &  16.7       &  -          \\ \hline
 \hline
\end{tabular}
}
\end{center}
\caption{Same as Table~\ref{tab:SUMMARY_enu} but for a Dirac dark matter particle.}\label{tab:SUMMARY_enu_DIRAC}
\end{table}

\newpage
\bibliographystyle{JHEP}
\bibliography{biblio}

\end{document}